\documentclass[]{fairmeta_gr2}

\title{ReasonRec: A Reasoning-Augmented Multimodal Agent for Unified Recommendation}

\author[1]{Yihua Zhang}
\author[1]{Mingfu Liang}
\author[1]{Jiyan Yang}
\author[1]{Rong Jin}
\author[1]{Wen-Yen Chen}
\author[1]{Yiping Han}
\author[1]{Huayu Li}
\author[1]{Buyun Zhang}
\author[1]{Liang Luo}
\author[1]{Frank Shyu}
\author[1]{Luke Simon}
\author[2]{Sijia Liu}
\author[3]{Tianlong Chen}
\author[1,\dagger]{Xi Liu}

\affiliation[1]{Meta AI}
\affiliation[2]{Michigan State University}
\affiliation[3]{The University of North Carolina at Chapel Hill}

\contribution[\dagger]{Project Tead}

\abstract{Recent advances in multimodal recommenders excel at feature fusion but remain opaque and inefficient decision-makers, lacking explicit reasoning and self-awareness of uncertainty. 
To address this, we introduce \textsc{{\ours}}, a reasoning-augmented multimodal agent structured around a three-stage explicit reasoning pipeline: \emph{Observe}, via a pretrained Vision-Language Model (VLM) encoder; \emph{Deliberate}, by formulating recommendation as chain-of-thought (CoT) reasoning tasks and explicitly quantifying prediction uncertainty through an evidence-horizon-aware curriculum; and \emph{Act}, through dynamic delegation of uncertain or challenging queries to lightweight classical recommendation models. 
Specifically, we propose a reasoning-aware visual instruction tuning strategy that systematically transforms diverse recommendation tasks into unified CoT prompts, enabling the VLM to explicitly articulate intermediate decision steps. Additionally, our evidence-horizon curriculum progressively enhances the reasoning complexity to better handle cold-start and long-tail user scenarios, significantly boosting model generalization. Furthermore, the uncertainty-guided delegation mechanism empowers the agent to assess its own confidence, strategically allocating computational resources to optimize both recommendation accuracy and inference efficiency. 
Comprehensive experiments on four standard recommendation tasks (sequential recommendation, direct recommendation, CTR prediction, and explanation generation) across five real-world datasets demonstrate that {\ours} achieves over 30\% relative improvement in key ranking metrics (\textit{e.g.}, HR@5, NDCG@5) compared to state-of-the-art multimodal recommenders. Crucially, {\ours} substantially reduces inference latency by dynamically delegating up to 35\% of queries to efficient sub-models without compromising accuracy. Extensive ablation studies further confirm that each proposed reasoning and planning mechanism individually contributes substantially to {\ours}'s overall effectiveness. Collectively, our results illustrate a clear pathway towards interpretable, adaptive, and efficient multimodal recommendation through explicit reasoning and agentic design.}

\date{\today}
\correspondence{\email{xliu1@meta.com}}
\metadata[Accepted to]{The 64th Annual Meeting of the Association for Computational Linguistics}

\usepackage{wrapfig}

\usepackage{multirow,mathtools }

\usepackage{pifont}
\usepackage{color, colortbl}

\usepackage{blindtext}
\usepackage{lipsum}

\usepackage{multirow}
\usepackage{listings}

\usepackage{bbm}

\usepackage [english]{babel}
\usepackage[autostyle, english = american]{csquotes}

\usepackage{pifont}
\usepackage{url}
\usepackage[most]{tcolorbox}

\usepackage{lipsum}
\usepackage{soul}
\usepackage{xcolor}
\usepackage{wrapfig}
\usepackage{multirow,mathtools } 

\usepackage{adjustbox}
\MakeOuterQuote{"}

\usepackage{microtype}
\usepackage{graphicx}
\usepackage{booktabs} 

\usepackage{hyperref}

\usepackage{amsmath}

\usepackage{nicefrac}       

\usepackage{tablefootnote}

\usepackage[table]{xcolor}
\usepackage{array}
\usepackage{amsfonts}
\usepackage{amssymb}
\usepackage{enumitem}

\usepackage{pifont}
\newcommand{\cmark}{\textcolor{green}{\ding{51}}}
\newcommand{\xmark}{\textcolor{red}{\ding{55}}}

\usepackage{color, colortbl}
\definecolor{Gray}{gray}{0.93}
\definecolor{Orange}{rgb}{1,0.5,0}
\definecolor{DGray}{gray}{0.83}
\definecolor{LightCyan}{rgb}{0.88,1,1}

\definecolor{WarnREd}{rgb}{1,0.4,0.4}
\definecolor{WarnOrange}{rgb}{1,0.682,0.502}
\definecolor{WarnPink}{rgb}{0.9176, 0.7215, 0.7215}
\definecolor{GoodGreen}{rgb}{0.5019, 0.9215, 0.6039}

\usepackage[T1]{fontenc}

\definecolor{styleblue}{HTML}{504099}
\definecolor{mypurple}{HTML}{9391ff}

\definecolor{bluegray}{rgb}{0.4, 0.6, 0.8}
\definecolor{ceruleanblue}{rgb}{0.16, 0.32, 0.75}

\hypersetup{
colorlinks=true,
citecolor=ceruleanblue,
linkcolor=ceruleanblue,
urlcolor=black
}

\usepackage{amsmath,amsfonts,bm}

\def\eqref#1{(\ref{#1})}

\def\1{\bm{1}}

\DeclareMathAlphabet{\mathsfit}{\encodingdefault}{\sfdefault}{m}{sl}
\SetMathAlphabet{\mathsfit}{bold}{\encodingdefault}{\sfdefault}{bx}{n}

\newcommand{\ours}{\textsc{ReasonRec}}

\begin{document}

\maketitle

\section{Introduction}
\label{sec: intro}

Recent advances in large language models (LLMs), exemplified by DeepSeek-R1~\cite{guo2025deepseek}, GPT-4o~\cite{hurst2024gpt}, and Gemini~\cite{team2023gemini}, have sparked a widespread ``reasoning renaissance'' in artificial intelligence research. These models leverage explicit chain-of-thought (CoT) ~\cite{wei2022chain} reasoning to achieve remarkable performance across various complex reasoning tasks, including logical inference~\cite{li2024crowdsourced}, mathematical problem-solving~\cite{AIME2024}, and planning~\cite{xie2024travelplanner}. 
Such reasoning capabilities not only improve model interpretability but also significantly enhance generalization and trustworthiness. Despite these successes, the power of explicit reasoning~\cite{guo2025deepseek} has yet to be systematically explored and exploited in multimodal recommendation~\cite{xie2024large}, a crucial domain inherently requiring nuanced and transparent decision-making.

\begin{figure}[t]
    \centering
    \includegraphics[width=0.7\linewidth]{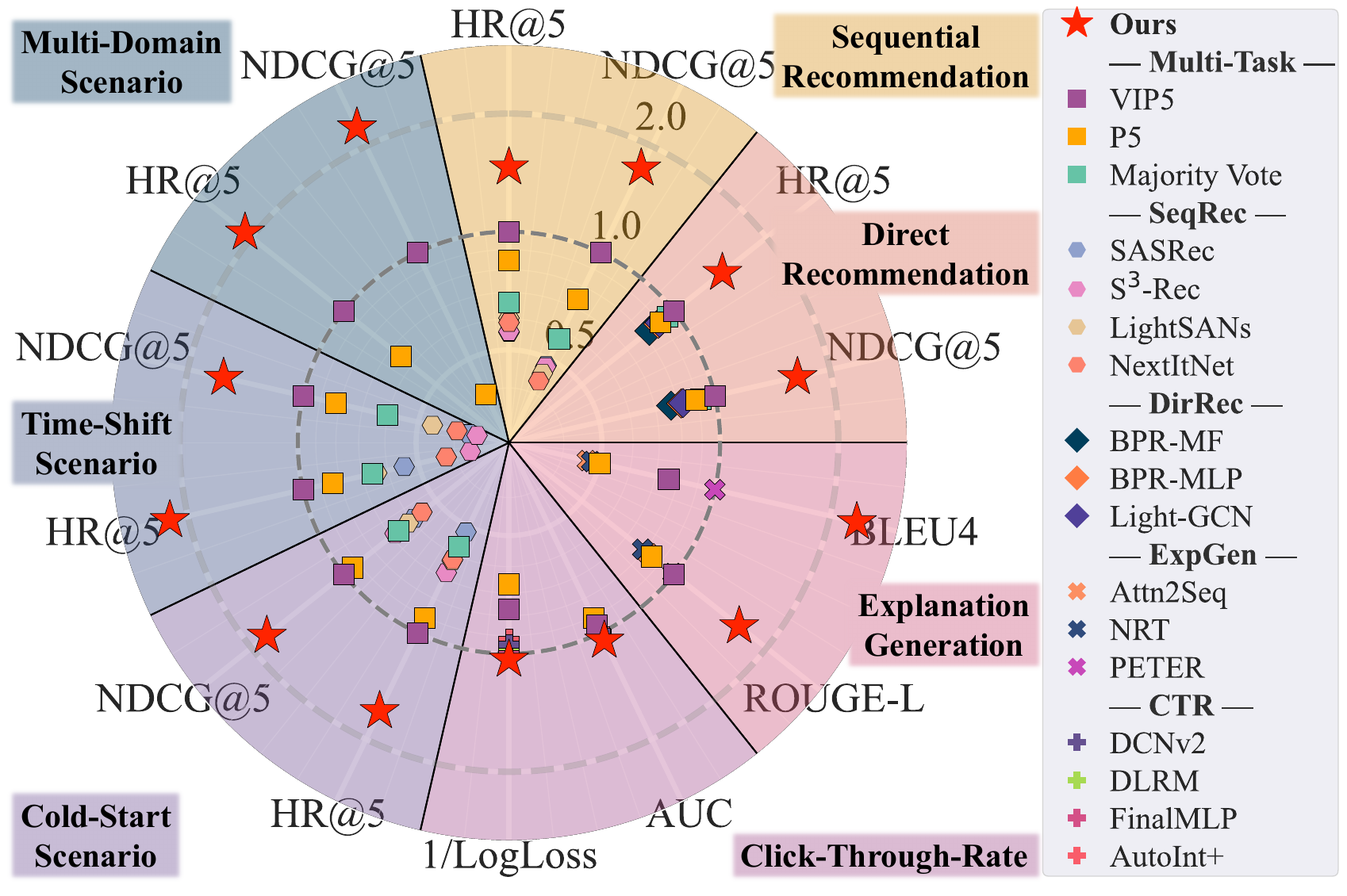}
    \vspace*{-0.5em}
    \caption{Performance comparison across four basic recommendation tasks and three challenging scenarios. Each task (color-coded region) is evaluated on two metrics. All results are normalized such that the {best baseline performance is set to $\mathbf{1.0}$} for each metric-task pair. Markers representing {\ours} indicate the relative improvement ratio over the strongest baseline, with values greater than demonstrating superiority. Experiment setups and results can be found in Sec.\,\ref{sec: exp}.}
    \label{fig: teaser}
    \vspace*{-1em}
\end{figure}

Multimodal recommendation~\cite{cheng2023image,geng2023vip5}fundamentally involves intricate multi-step deliberation processes. To effectively fulfill user needs, a recommender must first accurately infer user intent from sparse and ambiguous interactions~\cite{ytdnn_goog_recsys16,wdl_goog_dlrs16,dcnv2_www21}. Next, it must carefully evaluate the semantic content embedded in diverse multimodal signals, such as visual imagery and descriptive text, to estimate relevance. Finally, recommenders must explicitly reason about complex utility trade-offs, balancing user satisfaction, content freshness, and system efficiency~\cite{kang2018self,zhou2020s3,fan2021lighter,yuan2019simple}. However, existing multimodal recommendation systems typically rely on a singular, black-box forward inference, collapsing all these intricate decisions into an opaque scoring mechanism~\cite{cheng2023image,geng2023vip5}. This implicit approach severely limits transparency, interpretability, and robustness, especially in challenging scenarios like cold-start and cross-domain tasks~\cite{zhang2024wukong,afn,song2019autoint,dlrm}.

Most contemporary Vision-Language Models (VLMs)~\cite{zong2024safety,liu2023visual, liu2024improved, zhu2023minigpt, ye2023mplug, wang2023visionllm, li2023mimic, alayrac2022flamingo, awadalla2023openflamingo} in recommendation embed multimodal signals implicitly within their transformer architectures. While effective at feature fusion, these models fail to explicitly articulate their reasoning processes~\cite{zhang2024wukong}, resulting in recommendations that are hard to interpret and audit. More critically, these single-pass architectures lack the self-awareness to diagnose the uncertainty and thus cannot dynamically adapt by invoking specialized external knowledge or computationally efficient sub-models. Thus, existing multimodal recommenders suffer from poor reliability in low-confidence situations and waste computational resources by uniformly treating all recommendation decisions as equally complex.

Motivated by these limitations, we propose to fundamentally rethink the design of multimodal recommenders around an explicit reasoning paradigm. Specifically, we ask: Can we build an agentic VLM that reasons explicitly through chain-of-thought, transparently assesses its own uncertainty, and dynamically delegates uncertain or challenging cases to external, lightweight specialists? Addressing this question not only promises to significantly enhance the interpretability and generalization capabilities of multimodal recommenders but also introduces a novel, reasoning-driven agent framework into the broader recommendation landscape. In this work, we introduce ReasonRec, a reasoning-augmented multimodal recommendation agent that transforms recommendation into a transparent and adaptive decision process. Our key contributions include:

$\bullet$ A reasoning-aware instruction tuning framework that reformulates diverse recommendation tasks, including sequential recommendation, direct recommendation, CTR prediction, and explanation generation, into a unified chain-of-thought (CoT) format, enabling the VLM to verbalize intermediate reasoning steps and improve task alignment.

$\bullet$ An evidence-horizon curriculum learning strategy that gradually expands the complexity of reasoning chains by controlling user-item sparsity levels during training. This approach enhances generalization in cold-start and long-tail scenarios.

$\bullet$ An uncertainty-guided tool delegation mechanism that equips the agent with the ability to assess its the query risk and dynamically invoke lightweight classical models, balancing computational cost and predictive robustness.

$\bullet$ Extensive empirical validation across five public benchmarks and four recommendation tasks, demonstrating that ReasonRec achieves over 30\% improvement in HR@5 and NDCG@5 compared to prior state-of-the-art models; adapts effectively to cold-start, time-shift, and multi-domain scenarios; and reduces inference cost by more than 30\% through selective delegation.

\section{Related Work}
\label{sec: related_work}

\textbf{Generative models for recommendation systems.}
Recent advances in recommendation systems have witnessed a paradigm shift toward generative architectures and multimodal content understanding. Building upon the foundation of large language models (LLMs) \cite{li2023large}, pioneering works like Transformers4Rec \cite{de2021transformers4rec} and BERT4Rec \cite{sun2019bert4rec} employ Transformer architectures for sequential modeling, establishing frameworks for transferable recommendation through language model pretraining. Subsequent innovations extend this paradigm through diverse representation learning strategies: UniSRec \cite{hou2022towards} constructs item embeddings from descriptive texts rather than static IDs, while TransRec \cite{wang2022transrec} integrates multimodal user feedback through BERT and ResNet encoders for content-based personalization. The emergence of prompt engineering has further enriched this landscape, with PETER \cite{li2021personalized} and PEPLER \cite{li2022personalized} developing continuous prompt templates to encode user-item interactions while generating textual rationales for recommendations. Architectural unification efforts like M6-Rec \cite{cui2022m6} convert behavioral patterns into text sequences for Transformer processing, enabling task-adaptive fine-tuning through customized loss functions. P5 \cite{geng2022recommendation} and OpenP5 \cite{xu2023openp5} achieve cross-task generalization by implementing instruction-tuned LLMs that represent user-item relationships through natural language interfaces, later extended by P5-ID \cite{hua2023index} through novel item indexing schemes combining sequential, collaborative, and semantic signals. Multimodal generation techniques have simultaneously evolved across three directions: auxiliary feature integration, explainable recommendation, and semantic structure discovery. Early approaches like VBPR \cite{he2016vbpr} and PiNet \cite{meng2020heterogeneous} enhance collaborative filtering through visual feature extraction and heterogeneous preference modeling, respectively, while JRL \cite{zhang2017joint} pioneers joint multimodal representation learning. Domain-specific generators have emerged for fashion \cite{hou2019explainable,verma2020fashionist,chen2019personalized}, travel \cite{geng2022improving}, and culinary recommendations \cite{meng2020heterogeneous}, producing visually-grounded explanations. 

\noindent\textbf{Vision-language models and agents.} The evolution of Large Language Models has catalyzed next-generation vision-language architectures, transcending traditional visual-language systems through LLM-powered linguistic reasoning. Pioneered by architectures such as LLaVA-series~\cite{liu2023visual, liu2023improved, llava-rlhf, llavaplus}, BLIP-family~\cite{blip-2, instructblip}, and MiniGPT-4~\cite{minigpt4}, these models demonstrate exceptional visual dialog capabilities via LLM-based language encoders. However, their computational footprint, typically requiring 7B-65B parameters, creates deployment bottlenecks for edge/mobile platforms demanding real-time responsiveness. While proprietary models like Gemini~\cite{gemini} address this via scaled variants (\textit{e.g.}, 1.8B-parameter Nano for smartphones), their closed-source nature limits adaptability. Open-source initiatives like MobileVLM~\cite{chu2023mobilevlm} develop compact architectures (e.g., 2.7B-parameter mobileLLaMA) to bridge this gap. In this work, we for the first time exploit a visual instruction tuning framework for recommendation system for pretrained VLMs.

\section{ReasonRec: A Reasoning-Augmented Recommendation Agent}
\label{sec: method}

To enable explicit reasoning and adaptive decision-making in multimodal recommendation, we propose \textbf{ReasonRec}, a unified agentic framework structured as a three-stage reasoning pipeline: \emph{Observe $\rightarrow$ Deliberate $\rightarrow$ Act}. Next, we detail each of these stages. 

\subsection{Observer: Visual and Textual Perception}
\label{subsec: observer}

The first stage, the \emph{Observer}, extracts multimodal information critical for informed reasoning. Specifically, given user history representations $\mathcal{H}_u$ (past interactions) and candidate item information $\mathcal{I}$ (visual image $\mathbf{x}_v$, query text $\mathbf{x}_q$ and metadata), the Observer employs a pretrained Vision-Language Model (VLM) encoder~\cite{liu2023visual,liu2023improved} to generate a unified embedding $\mathbf{h} = \text{VLMEncoder}(\mathbf{x}_v, \mathbf{x}_q, \mathcal{H}_u)$.
These embeddings provide a rich representation capturing user intent and item semantics, thus forming a robust foundation for explicit reasoning.

\subsection{Deliberator: Explicit Reasoning and Self-Reflection}
\label{subsec: deliberator}

The \emph{Deliberator} explicitly performs reasoning by reformulating recommendation tasks into structured instruction-following problems. Traditional VLMs lack structured interaction modeling and task alignment, thus failing in sparse interaction scenarios. To mitigate these issues, we introduce \textbf{Reasoning-Aware Visual Instruction Tuning (R-VIT)}, comprising three critical innovations:

\textbf{(1) Task formulation as VQA.} We convert recommendation tasks into structured vision-question-answering (VQA) instructions. Formally, given multimodal inputs, the Deliberator generates outputs in a structured prompt-response format:

\vspace*{-1.5em}
{\small
\begin{align}
\texttt{User}:  \mathbf x_v \,<\!\backslash n\!>\, \mathbf x_q \texttt{<STOP>}\; \nonumber\\
    \texttt{Assistant}: [\text{Thought Tokens}] \rightarrow \mathbf{y}\,\texttt{<STOP>}, \nonumber 
\label{eq: rec_as_vqa}
\vspace*{-2em}
\end{align}
}%

\noindent where `[Thought Tokens]` explicitly verbalize intermediate reasoning steps (\textbf{Fig.\,\ref{fig: instruction_data_example}}). Training optimizes an autoregressive objective to ensure reasoning consistency.

\textbf{(2) Template mixtures for generalization.} We use \textit{multiple diverse instruction templates per task} to prevent overfitting and improve generalization to unseen variations. Instead of a fixed format, we construct multiple templates to express similar semantic in different linguistic styles and structures. This encourages the model to focus on \textit{task semantics rather than patterns}, improving robustness against distribution shifts. As shown in Fig.\,\ref{fig: ablation}, this mixture-of-templates approach significantly enhances training effectiveness.

\begin{figure}[t]
    \centering
    \begin{tabular}{cc}
       \hspace*{-5mm}\includegraphics[width=0.3\linewidth]{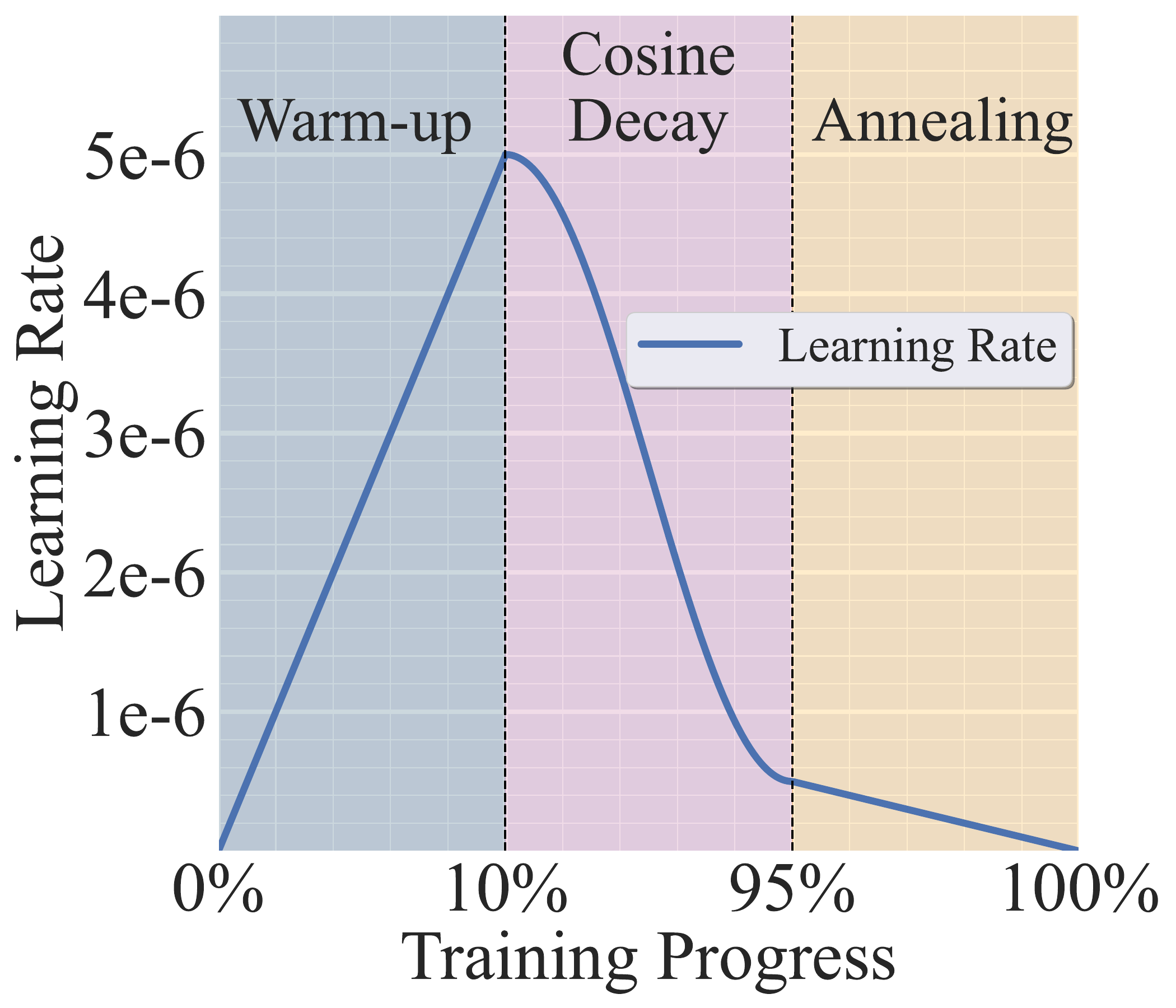}  
       & \hspace*{-5mm}  \includegraphics[width=0.3\linewidth]{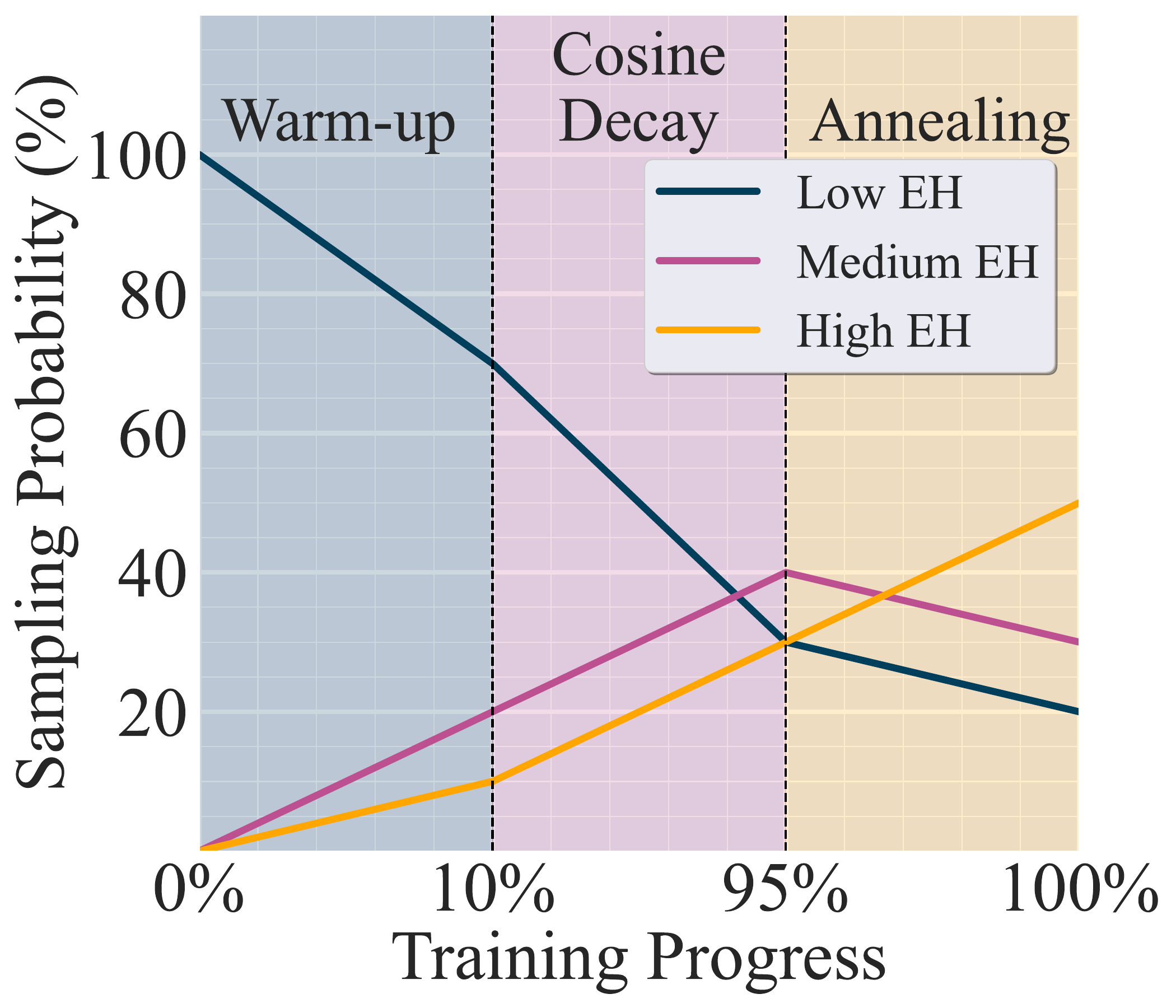} \vspace*{-0.5em}\\
       \hspace*{-2mm} \footnotesize{(A) Learning rate schedule.} 
       & \hspace*{-5mm} \footnotesize{(B) EH sampling.}
    \end{tabular}
    \vspace*{-0.5em}
    \caption{
Evidence-horizon curriculum learning. Users are grouped into \emph{Low}, \emph{Medium}, and \emph{High EH} based on interaction sparsity. The learning-rate schedule (A) is aligned with the sampling policy (B): training begins with dense users, gradually shifts to harder cases, and finally focuses on high-EH (cold-start) users. This staged strategy enhances ReasonRec's reasoning under sparsity while mitigating forgetting on warm-user patterns.
}
    \label{fig: coldness_aware_sampling}
    \vspace*{-1em}
\end{figure}

\textbf{(3) Evidence horizon quantification.} A major challenge in recommender systems is handling data sparsity, particularly in cold-start scenarios where users or items lack sufficient historical interactions. Traditional approaches struggle under these conditions, as models often rely heavily on frequent patterns while failing to capture long-tail behaviors. To address this issue, we introduce an \textbf{evidence-horizon-aware curriculum learning} strategy, gradually adapting the VLM to varying levels of data sparsity. This involves (i) defining a formal evidence horizon metric to characterize user-item sparsity and (ii) progressively adjusting the difficulty of training samples over time.
To measure data sparsity, we define an \textit{evidence horizon score} $C(u)$ for a user $u$ based on their historical interactions $\displaystyle C(u) = 1 - \frac{|\mathcal{I}_u|}{\max_{u'} |\mathcal{I}_{u'}|}$, 
where $|\mathcal{I}_u|$ denotes the interaction count of user $u$, and $\max_{u'} |\mathcal{I}_{u'}|$ is the maximum interaction count among all users. A higher evidence horizon score indicates fewer interactions, representing greater recommendation difficulty. During training, we explicitly incorporate this evidence horizon score into instruction templates, guiding the VLM to adjust its reasoning complexity based on the query difficulty. As demonstrated later, this explicit evidence horizon modeling significantly enhances the model’s uncertainty management and adaptive tool utilization.

\textbf{(4) Evidence-horizon-aware curriculum learning.} Rather than uniformly sampling all training data, we progressively increase the difficulty of training samples, inspired by data-mixing techniques in LLM pretraining~\cite{dubey2024llama}. Specifically, the training consists of three distinct phases (\textbf{Fig.\,\ref{fig: coldness_aware_sampling}}): \ding{172} Warm-up Phase: Initially, the model learns from users with abundant historical interactions (low evidence horizon), capturing strong user-item correlations; \ding{173} Progressive Learning Phase: Gradually introduces medium evidence horizon users and items, improving generalization to sparser distributions; \ding{174} Cold-start Emphasis Phase: Intensively trains on high evidence horizon scenarios with reduced learning rates~\cite{dubey2024llama,liu2024deepseek}, enhancing robustness against challenging, sparse data conditions.


\subsection{Actuator: Uncertainty-Guided Tool Delegation}
\label{subsec: actuator}


\noindent The \emph{Actuator} dynamically decides whether to delegate the reasoning task to lightweight classical models (tools) or rely directly on the VLM, guided by both uncertainty and evidence horizon. Below we detail the mechanisms enabling adaptive, efficient decision-making:

\begin{figure}[t]
    \centering
    \includegraphics[width=0.5\linewidth]{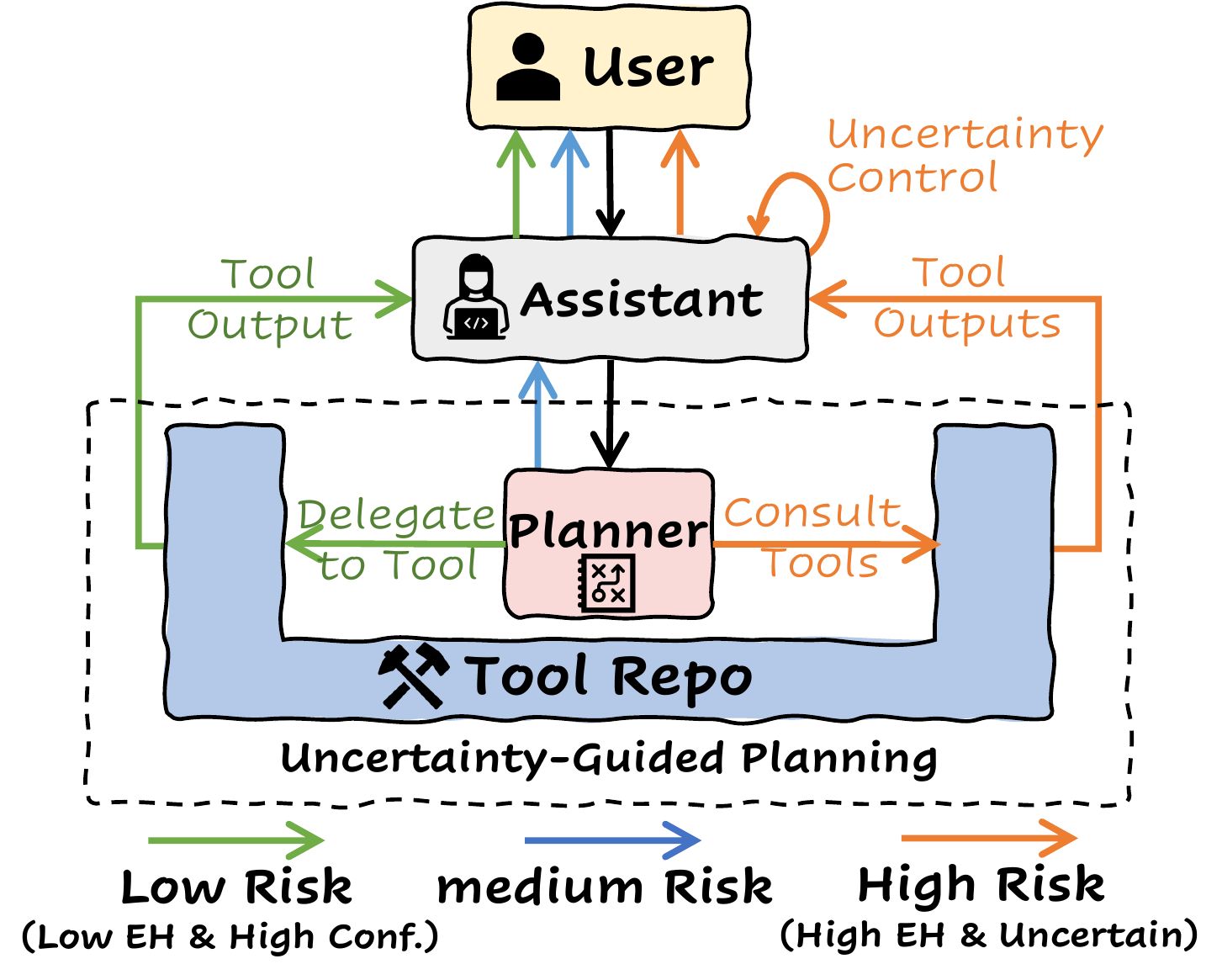}
    \vspace*{-1em}
    \caption{ Overview of ReasonRec's inference pipeline with \emph{uncertainty-guided planning}. The planner jointly considers a user’s {evidence horizon} (EH) and model {confidence} to assign each query to one of three risk levels. \emph{Low-risk} queries (low EH, high confidence) are \emph{delegated} to lightweight models from the \emph{tool repository}; \emph{medium-risk} queries are handled directly by the VLM; \emph{high-risk} queries (high EH, low confidence) \emph{consult tools} to refine the decision.
    }
    \vspace*{-1em}
    \label{fig: planning}
\end{figure}

\noindent\textbf{Uncertainty-guided planning with classical models.} Deploying large-scale VLMs for recommendation tasks requires balancing computational efficiency with predictive accuracy. While VLMs exhibit powerful reasoning, their inference costs are substantially higher than classical recommendation methods. To resolve this, we propose a \textbf{risk-aware planning mechanism}, dynamically choosing between VLM and lightweight classical models based on evidence horizon and model confidence.
Specifically, as illustrated in \textbf{Fig.\,\ref{fig: planning}}, we categorize queries according to their risk level: \ding{172} \textbf{Low-Risk}: Users with low evidence horizon and high model confidence - delegated directly to classical recommendation models for efficient inference. \ding{173} \textbf{Medium-Risk}: Moderate uncertainty queries handled directly by the VLM, exploiting internal reasoning capacities. \ding{174} \textbf{High-Risk}: Users with high evidence horizon and significant uncertainty, invoking multiple classical models whose aggregated outputs refine subsequent VLM-based reasoning. This hierarchical approach balances computational cost and accuracy, leveraging the VLM’s reasoning prowess specifically for complex cases.

\noindent\textbf{Tool repository.} To ensure efficiency at scale, we maintain a \textbf{tool repository} of classical models, including matrix factorization, graph-based models, and two-tower architectures (\textit{e.g.}, xDeepFM) optimized for efficient CTR prediction. These tools directly handle low-risk cases, significantly reducing VLM inference overhead without accuracy loss.

\noindent\textbf{Uncertainty-aware inference and tool integration.} In high-risk scenarios, the Deliberator initially predicts with explicit uncertainty (\textit{e.g.}, ``Recommend Item Y. Confidence: 0.53''). Subsequently, classical tools from the repository refine this initial prediction, improving final accuracy. 

In summary, the integrated reasoning pipeline (\emph{Observe $\rightarrow$ Deliberate $\rightarrow$ Act}), combined with explicit uncertainty estimation, adaptive tool delegation, and evidence-horizon-aware curriculum learning, empowers ReasonRec with interpretability, robustness, and efficiency, particularly in challenging multimodal recommendation scenarios.

\begin{table*}[t]
    \centering
    \caption{Performance comparison in {sequential and direct recommendation} using Amazon and Pixel-1M datasets.}
    \vspace*{-0.5em}
    \resizebox{.85\textwidth}{!}{%
    \begin{tabular}{l|cc|cc|cc|cc|cc}
        \toprule[1pt]
        \midrule
        \multirow{2}{*}{\textbf{Methods}} & 
        \multicolumn{2}{c|}{\textbf{Sports}} & 
        \multicolumn{2}{c|}{\textbf{Beauty}} & 
        \multicolumn{2}{c|}{\textbf{Clothing}} & 
        \multicolumn{2}{c|}{\textbf{Toys}} &
        \multicolumn{2}{c}{\textbf{Pixel-1M}} \\
        \cmidrule(lr){2-3} \cmidrule(lr){4-5} \cmidrule(lr){6-7} \cmidrule(lr){8-9} \cmidrule(lr){10-11}
         & HR@5 & NDCG@5 & HR@5 & NDCG@5 & HR@5 & NDCG@5 & HR@5 & NDCG@5 & HR@5 & NDCG@5 \\
        \midrule
        \multicolumn{11}{c}{\textbf{Sequential Recommendation}} \\
        \midrule
        SASRec    & 0.0289 & 0.0175 & 0.0403 & 0.0297 & 0.0132 & 0.0126 & 0.0463 & 0.0338 & 0.0116 & 0.0107 \\
        S$^3$-Rec  & 0.0274 & 0.0189 & 0.0415 & 0.0286 & 0.0110 & 0.0105 & 0.0443 & 0.0344 & 0.0168 & 0.0111 \\
        LightSANs & 0.0260 & 0.0170 & 0.0435 & 0.0250 & 0.0185 & 0.0109 & 0.0481 & 0.0354 & 0.0165 & 0.0108 \\
        NextItNet & 0.0258 & 0.0197 & 0.0427 & 0.0248 & 0.0192 & 0.0072 & 0.0470 & 0.0327 & 0.0140 & 0.0099 \\
        Majority Vote & 0.0313    & 0.0211    & 0.0492    & 0.0315    & 0.0214    & 0.0199    & 0.0512    & 0.0382    & 0.0144    & 0.0135    \\
\midrule
        P5        & 0.0275 & 0.0176 & 0.0483 & 0.0398 & 0.0499 & 0.0392 & 0.0694 & 0.0523 & 0.0188 & 0.0115 \\
        VIP5      & 0.0436 & 0.0371 & 0.0565 & 0.0489 & 0.0623 & 0.0597 & 0.0712 & 0.0596 & 0.0197 & 0.0123 \\
        UniMP     & 0.0515 & 0.0419 & 0.0602 & 0.0531 & 0.0679 & 0.0632 & 0.0794 & 0.0647 & 0.0256 & 0.0170 \\
\midrule
        {\ours}      & \textbf{0.0721} & \textbf{0.0694} & 
                    \textbf{0.0797} & \textbf{0.0944} & 
                    \textbf{0.0832} & \textbf{0.1011} & 
                    \textbf{0.1032} & \textbf{0.0901} & 
                    \textbf{0.0315} & \textbf{0.0218} \\
\midrule
        \multicolumn{11}{c}{\textbf{Direct Recommendation}} \\
\midrule
        BPR-MF    & 0.1478 & 0.0897 & 0.1426 & 0.0913 & 0.1280 & 0.0735 & 0.1023 & 0.0641 & 0.0356 & 0.0231 \\
        BPR-MLP   & 0.1592 & 0.0945 & 0.1381 & 0.0891 & 0.1421 & 0.0822 & 0.1171 & 0.0721 & 0.0384 & 0.0253 \\
        Light-GCN & 0.1549 & 0.0911 & 0.1501 & 0.0904 & 0.1475 & 0.0831 & 0.1121 & 0.0744 & 0.0362 & 0.0281 \\
        Majority Vote & 0.1614    & 0.0993    & 0.1612    & 0.1043    & 0.1514    & 0.0931    & 0.1229    & 0.0813    & 0.0403 & 0.0299 \\
\midrule
        P5   & 0.1583 & 0.1132 & 0.1681 & 0.1123 & 0.1001 & 0.0639 & 0.1232 & 0.0841 & 0.0539 & 0.0243 \\
        VIP5 & 0.1791 & 0.1241 & 0.1739 & 0.1113 & 0.1299 & 0.0871 & 0.1245 & 0.0829 & 0.0624 & 0.0392 \\
        UniMP & 0.1940 & 0.1372 & 0.1825 & 0.1220 & 0.1378 & 0.0965 & 0.1302 & 0.0887 & 0.0703 & 0.0451 \\
\midrule
        {\ours}      & \textbf{0.2439} & \textbf{0.1732} & 
                    \textbf{0.2351} & \textbf{0.1655} & 
                    \textbf{0.1998} & \textbf{0.1523} & 
                    \textbf{0.1839} & \textbf{0.1671} & 
                    \textbf{0.0991} & \textbf{0.0725} \\
        \bottomrule[1pt]
    \end{tabular}}
    \vspace*{-1em}
    \label{tab: sequential_and_direct_recommendation}
\end{table*}

\section{Experiments}
\label{sec: exp}

We provide a comprehensive evaluation on the proposed {\ours} in four commonly-used recommendation tasks. We also consider three challenging recommendation settings, including \textit{cold-start}, \textit{time-shift}, and \textit{multi-domain} scenarios. Abundant ablation studies are also conducted to demonstrate the effectiveness of our proposed planning strategy.

\subsection{Experiment Setups}

\textbf{Model and datasets.}  
We evaluate {\ours} across four key recommendation tasks: sequential recommendation, direct recommendation, explanation generation, and CTR prediction, using LLaVA1.5-7B~\citep{liu2023improved} as the underlying VLM. \textbf{Amazon Review Dataset:} We use four categories from the Amazon Review dataset: \textit{Clothing, Shoes \& Jewelry}, \textit{Sports \& Outdoors}, \textit{Beauty}, and \textit{Toys \& Games}. Each includes user purchase histories, item metadata, textual reviews, and images (see Tab.~\ref{tab: amazon_review_statistics}). \textbf{Pixel-1M Dataset~\cite{cheng2023image}:} This large-scale image-centric dataset contains over 1M users, 100K images, and 20M user--image interactions. Unlike ID-based datasets, it enables learning directly from raw image pixels. We adopt a leave-one-out split: the last interaction per user for testing, the second-to-last for validation, and the rest for training.

\noindent\textbf{Baselines compared in each task.}  
We compare {\ours} against stateful baselines for each task:  
$\bullet$ \textbf{Sequential Recommendation (Tab.~\ref{tab: sequential_and_direct_recommendation}):} We include four classical sequential models: SASRec~\cite{kang2018self}, S$^3$-Rec~\cite{zhou2020s3}, LightSANs~\cite{fan2021lighter}, and NextItNet~\cite{yuan2019simple}, and three generative baselines: P5~\cite{geng2022recommendation}, VIP5~\cite{geng2023vip5}, and UniMP~\cite{wei2024towards}. Following PixelRec~\cite{cheng2023image}, we replace item ID embeddings in classical models with visual features to ensure a fair multimodal comparison.  
$\bullet$ \textbf{Direct Recommendation (Tab.~\ref{tab: sequential_and_direct_recommendation}):} Baselines include classical models BPR-MF~\cite{rendle2012bpr}, BPR-MLP~\cite{rendle2012bpr}, LightGCN~\cite{he2020lightgcn}, and the same generative baselines used above.  
$\bullet$ \textbf{Explanation Generation (Tab.~\ref{tab: amazon_explanation}):} We follow VIP5~\cite{geng2023vip5} and compare with Attn2Seq~\cite{dong2017learning}, NRT~\cite{dong2017learning}, PETER~\cite{li2021personalized}, as well as P5 and VIP5.  
$\bullet$ \textbf{CTR Prediction (Tab.~\ref{tab: click_through_rate}):} Following~\cite{zhang2024wukong}, we adopt AFN+~\cite{afn}, AutoInt+~\cite{song2019autoint}, DLRM~\cite{dlrm}, DCNv2~\cite{wang2021dcn}, FinalMLP~\cite{mao2023finalmlp}, MaskNet~\cite{wang2021masknet}, and xDeepFM~\cite{lian2018xdeepfm}, implemented via the BARS evaluation framework~\cite{bars,bars2}.  In sequential, direct, and CTR tasks, we additionally report a ``majority vote'' baseline that averages predictions from all classical models. This helps isolate the effect of our planner and clarify that the performance of {\ours} is not simply due to tool usage.

\noindent\textbf{Training and evaluation setups.} Non-generative models are trained separately for dataset. Generative models are trained per dataset using mixed-task instruction tuning. For the multi-domain challenge (across Sports, Beauty, Clothing, and Toys), P5, VIP5, UniMP and {\ours} are trained across domains. More details are provided in Appx.~\ref{app: exp_setup}.
For sequential and direct recommendation, we report \textbf{HR@5} and \textbf{NDCG@5}. For explanation generation, we use \textbf{BLEU4} and \textbf{ROUGEL}. For CTR prediction, we use AUC and LogLoss, where higher AUC and lower LogLoss are preferred.

\begin{table*}[t]
\centering
\caption{Performance comparison in click-through-rate task using Amazon Review and Pixel-1M datasets.}
\vspace*{-0.5em}
\resizebox{0.8\textwidth}{!}{%
\begin{tabular}{l|cc|cc|cc|cc|cc}
    \toprule[1pt]
\midrule
    \multirow{2}{*}{\textbf{Methods}} & 
    \multicolumn{2}{c|}{\textbf{Sports}} & 
    \multicolumn{2}{c|}{\textbf{Beauty}} & 
    \multicolumn{2}{c|}{\textbf{Clothing}} & 
    \multicolumn{2}{c|}{\textbf{Toys}} &
    \multicolumn{2}{c}{\textbf{Pixel-1M}} \\
    \cmidrule(lr){2-3} \cmidrule(lr){4-5} \cmidrule(lr){6-7} \cmidrule(lr){8-9} \cmidrule(lr){10-11}
     & \textbf{AUC} ($\uparrow$) & \textbf{LogLoss} ($\downarrow$) & \textbf{AUC} ($\uparrow$) & \textbf{LogLoss} ($\downarrow$) & \textbf{AUC} ($\uparrow$) & \textbf{LogLoss} ($\downarrow$) & \textbf{AUC} ($\uparrow$) & \textbf{LogLoss} ($\downarrow$) & \textbf{AUC} ($\uparrow$) & \textbf{LogLoss} ($\downarrow$) \\
\midrule
     AutoInt+       & 0.8033 & 0.2432 & 0.8732 & 0.2031 & 0.7224 & 0.3296 & 0.7533 & 0.3115 & 0.5123 & 0.4993 \\
     DLRM           & 0.8145 & 0.2533 & 0.8819 & 0.1672 & 0.7174 & 0.3442 & 0.7542 & 0.3411 & 0.5472 & 0.4173 \\
     FinalMLP       & 0.7984 & 0.2411 & 0.8801 & 0.1993 & 0.7253 & 0.3213 & 0.7635 & 0.3021 & 0.5524 & 0.3984 \\
     DCNv2          & 0.8024 & 0.2395 & 0.8825 & 0.1742 & 0.7297 & 0.3459 & 0.7513 & 0.3571 & 0.5323 & 0.4242 \\
\midrule
    P5              & 0.6532 & 0.4473 & 0.7744 & 0.4336 & 0.6743 & 0.4544 & 0.6931 & 0.4946 & 0.5832 & 0.3672  \\
    VIP5            & 0.6812 & 0.3985 & 0.8415 & 0.3573 & 0.7025 & 0.4135 & 0.7113 & 0.3846 & 0.6031 & 0.3449  \\
\midrule
    {\ours}       & \textbf{0.8429} & \textbf{0.2445} & 
                \textbf{0.9113} & \textbf{0.1993} & 
                \textbf{0.7443} & \textbf{0.3139} & 
                \textbf{0.7815} & \textbf{0.3329} & 
                \textbf{0.6311} & \textbf{0.3222} \\
\midrule
    \bottomrule[1pt]
\end{tabular}}
\label{tab: click_through_rate}
\vspace*{-1em}
\end{table*}

\subsection{Experiment Results}

\textbf{A holistic comparison on sequential and direct recommendation tasks.} 
Tab.\,\ref{tab: sequential_and_direct_recommendation} demonstrates the state-of-the-art performance of {\ours} across both recommendation tasks. In sequential recommendation, \textit{first}, {\ours} achieves superior HR@5 and NDCG@5 across all datasets, surpassing individual baselines and the ``majority vote'' ensemble. This confirms that the gains stem from our multimodal instruction tuning and planning, rather than simple model combination. \textit{Second}, classical methods (SASRec, S$^3$-Rec, LightSANs, etc.) and their ensemble provide limited improvements, lagging significantly behind generative models. \textit{Third}, generative models (P5, VIP5) outperform classical ones but show diminished advantages on the large dataset Pixel-1M. In contrast, {\ours} consistently achieves strong performance and effectively captures sequential preferences. Direct recommendation focuses on static user--item interactions. \textit{First}, {\ours} again consistently leads in all metrics and datasets, particularly on Pixel-1M, indicating robustness to diverse user feedback. \textit{Second}, classical methods and their ensemble achieve moderate improvements but remain inferior to multimodal methods. \textit{Third}, while P5 and VIP5 show promising results on smaller Amazon datasets, they weaken on Pixel-1M. {\ours} effectively addresses these challenges via planning and reasoning.

\noindent\textbf{Performance on the explanation generation task.}
In Tab.~\ref{tab: amazon_explanation}, we evaluate textual explanation quality using BLEU4 and ROUGEL. \textit{First}, {\ours} achieves substantially higher scores across all domains, highlighting its superior capability in generating accurate explanations. \textit{Second}, multimodal baselines (P5, VIP5) outperform conventional methods (Attn2Seq, NRT, PETER) but remain behind {\ours}, confirming the benefit of instruction tuning in a single VLM agent. \textit{Third}, our planning mechanism effectively captures detailed item attributes and user reasoning, enabling more coherent and informative explanations aligned with user and product specifics.

\begin{table}[t]
    \centering
    \caption{Performance on sequential recommendation in cold-start scenarios on Pixel-1M dataset.}
    \vspace*{-0.5em}
    \resizebox{0.45\linewidth}{!}{%
    \begin{tabular}{lcccc}
        \toprule
        \multirow{2}{*}{\textbf{Method}} & 
        \multirow{2}{*}{\textbf{Metric}} & 
        \multicolumn{3}{c}{\textbf{Cold-start Level}} \\
        \cmidrule(lr){3-5}
        & & Normal & Medium & Coldest\\
        \midrule
        
        \multirow{2}{*}{SASRec} 
        & HR@5 & 0.0116 & 0.0079 & 0.0058 \\
        & NDCG@5 & 0.0107 & 0.0043 & 0.0026\\
        
        \multirow{2}{*}{LightSANs}
        & HR@5 & 0.0165 & 0.0087 & 0.0053 \\
        & NDCG@5 & 0.0108 & 0.0045 & 0.0031 \\
        
        \multirow{2}{*}{NextItNet}
        & HR@5 & 0.0140 & 0.0095 & 0.0068 \\
        & NDCG@5 & 0.0099 & 0.0046 & 0.0021 \\
        
        \multirow{2}{*}{Majority Vote}
        & HR@5 & 0.0144 & 0.0091 & 0.0043 \\
        & NDCG@5 & 0.0135 & 0.0041 & 0.0022 \\
        
        \midrule
        \multirow{2}{*}{P5}
        & HR@5 & 0.0173 & 0.0188 & 0.0082 \\
        & NDCG@5 & 0.0115 & 0.0095 & 0.0070 \\
        
        \multirow{2}{*}{VIP5}
        & HR@5 & 0.0184 & 0.0197 & 0.0107 \\
        & NDCG@5 & 0.0123 & 0.0094 & 0.0081 \\

        \multirow{2}{*}{UniMP}
        & HR@5 & 0.0224 & 0.0192 & 0.0155 \\
        & NDCG@5 & 0.0183 & 0.0114 & 0.0099 \\
        
        \midrule
        \multirow{3}{*}{{\ours}}
        & HR@5 & \textbf{0.0315} & \textbf{0.0283} & \textbf{0.0214} \\
        & NDCG@5 & \textbf{0.0218} & \textbf{0.0169} & \textbf{0.0144} \\
        & Tool Use Rate & $7.3\%$ & $13.2\%$ & $35.7\%$ \\
        \bottomrule[1pt]
    \end{tabular}}
    \label{tab: cold_start_performance}
    \vspace*{-1em}
\end{table}

\noindent \textbf{Performance on the CTR task.}
In Tab.~\ref{tab: click_through_rate}, {\ours} achieves the highest AUC and lowest LogLoss across all datasets, outperforming both classical and multimodal baselines. Classical CTR models (AutoInt+, DLRM, FinalMLP, DCNv2) show reasonable accuracy but lack precision in fine-grained click probability estimation. 
Generative methods (P5, VIP5), though effective in recommendation and explanation, struggle to distinguish subtle click signals, highlighting the challenge of adapting language models for binary prediction. {\ours} excels by combining multimodal reasoning with robust CTR, demonstrating that explicit planning enhances probability estimates.

\noindent\textbf{Robustness against cold recommendation setting.}
We partition the Pixel-1M test set into ten user groups based on training frequency: from \textit{Group 1} (cold-start users) to \textit{Group 10} (warm users), each containing 20,000 interactions. We report HR@5 and NDCG@5 over three aggregated splits: \textit{Coldest} (Groups 1--3), \textit{Medium} (Groups 4--6), and \textit{Normal} (Groups 1--10). As shown in Tab.~\ref{tab: cold_start_performance}, {\ours} consistently achieves top performance, indicating strong robustness to data sparsity. \textit{First}, classical sequential methods (SASRec, S$^3$-Rec, LightSANs, NextItNet) degrade rapidly as data becomes sparse, whereas {\ours} maintains significantly higher metrics. \textit{Second}, multimodal approaches (P5, VIP5) improve moderately in medium-sparsity scenarios but perform poorly in the coldest groups. \textit{Third}, {\ours} adaptively increases external tool usage from $7.3\%$ (warm scenarios) to $35.7\%$ (coldest scenarios), strategically delegating simpler tasks and leveraging advanced planning for sparse interactions. This adaptive capability underscores the effectiveness of our framework for cold-start recommendation.

\noindent\textbf{More challenging scenarios.} We further investigated the performance of {\ours} in three challenging scenarios, namely the recommendation system with time shift and multi-domain recommendation task. Due to the page limit, we report a normalized performance overview in \textbf{Fig.\,\ref{fig: teaser}} and more results can be found in Appx.\,\ref{app: exp_results}.

\subsection{Diving into Planning Mechanism: Effectiveness and Efficiency}

\noindent To confirm that {\ours}'s gains are not due to ensembling alone, we compare it against two strong baselines: \textbf{(A) Tool Ensemble Only}, which averages predictions from lightweight models without VLM or planning; and \textbf{(B) VLM Only (No Planning)}, a fine-tuned VLM applied uniformly to all queries without delegation. As shown in Tab.~\ref{tab: ablation_ensemble}, {\ours} consistently outperforms both in HR@5, achieving a 0.0315 relative gain over the best baseline. Compared to (A), our selective delegation based on uncertainty and evidence horizon yields more reliable predictions than naive ensembling. Compared to (B), {\ours} avoids inefficient overuse of the VLM on low-risk queries while maintaining robustness in high-risk cases. These results confirm the gains are from \emph{risk-aware planning}, not model aggregation.

\noindent\textbf{Task-wise tool utilization comparison under risk-aware planning.}
Since {\ours}'s planner dynamically delegates queries based on estimated risk, analyzing \emph{how often external tools are invoked} offers insight into its scheduling behavior, especially under high-risk conditions requiring both VLM and tool collaboration. In Tab.~\ref{tab:tool-usage}, we report the \emph{tool usage rate}, \textit{i.e.}, the percentage of test queries triggering at least one external model, across three tasks (sequential recommendation, direct recommendation, and CTR prediction) and five datasets. Two trends emerge. \textit{First}, usage rates vary most across datasets: Pixel-1M, with the highest sparsity and evidence horizon, sees the most frequent tool delegation (up to 34.1\% in CTR), indicating strong planner sensitivity to data uncertainty. \textit{Second}, while task type impacts usage slightly, the effect is smaller and less consistent, suggesting that \emph{{\ours}'s delegation is driven more by input-level risk than task structure}. These results affirm the planner's role in identifying high EH queries and selectively allocating computation to optimize performance-efficiency trade-offs.

\begin{table}[t]
\centering
\caption{
Analysis of {\ours} vs. tool outputs on Pixel-1M (Sequential Recommendation).
We report tool accuracy, {\ours} accuracy, the disagreement rate between {\ours} and tools, and {\ours}’s accuracy in disagreement cases.
}
\vspace*{-1em}
\resizebox{0.5\linewidth}{!}{
\begin{tabular}{lc}
\toprule
\textbf{Metric} & \textbf{Value} \\
\midrule
Tool accuracy & 58.4\% \\
{\ours} accuracy       & 69.3\% \\
Disagreement rate ({\ours} $\neq$ Tool) & 13.6\% \\
{\ours} accuracy on disagreement cases  & 84.2\% \\
\bottomrule
\end{tabular}}
\vspace*{-1em}
\label{tab:tool_vlm_conflict}
\end{table}

\noindent\textbf{Conflict resolution between tool outputs and reasoning.}
To examine how {\ours} handles conflicts between tools and its own reasoning, we analyze its behavior on the Pixel-1M (sequential recommendation). Specifically, we evaluate whether {\ours} blindly follows tools. We report four metrics: (1) standalone accuracy of the tool ensemble, (2) overall accuracy of {\ours}, (3) the proportion of predictions where {\ours} disagrees with the tools, and (4) {\ours}’s accuracy in disagreements. Tab.~\ref{tab:tool_vlm_conflict} shows tools alone achieve 58.4\%, while {\ours} improves to 69.3\%. Disagreements occur in 13.6\% of cases, and {\ours} reaches 84.2\% in these. This indicates that {\ours} does not passively follow tools, but overrides them when its reasoning offers better alternatives, since we add occasional noisy tool outputs during training, encouraging the model to treat tools as useful but non-authoritative.

\begin{wraptable}{r}{0.5\linewidth}
\centering
\vspace*{-1em}
\caption{
Comparison on accuracy and inference efficiency.
HR@5 metric and average inference time per query are reported.
}
\vspace*{-1em}
\resizebox{0.6\linewidth}{!}{
\begin{tabular}{lccc}
\toprule
\textbf{Method} & \textbf{HR@5} & \textbf{Time (ms)} \\
\midrule
SASRec  & 0.0116 & \textbf{143} \\
VIP5    & 0.0197 & 470  \\
{\ours} & \textbf{0.0315} & 499 \\
\bottomrule
\end{tabular}}
\vspace*{-1em}
\label{tab: efficiency_baseline}
\end{wraptable}

\noindent\textbf{{\ours} strikes a balance between accuracy and efficiency.} 
While VLMs are often criticized for high inference cost, {\ours} is a structured system that achieves a strong trade-off via planning and adaptive delegation. 
Tab.~\ref{tab: efficiency_baseline} shows {\ours} attains near-optimal accuracy (HR@5 = 0.0315), with only slightly higher latency than non-VLM systems like VIP5 (499\,ms vs. 470\,ms), and significantly better accuracy (0.0315 vs. 0.0197). 
It further achieves SOTA results across tasks, especially in cold-start settings where existing methods often fall short.
Critically, the planning component of {\ours} is lightweight. While Fig.~\ref{fig: planning} shows human-readable natural language for clarity, the actual reasoning traces are in more structured way (e.g., ``Confidence: 0.532; Tools: A/B'') and brief CoT-style checks. These require only shallow decoding with negligible overhead. In low-risk cases, the system bypasses the VLM and adopt the results directly from the tool, keeping inference cost well within practical limits.

\begin{table}[t]
    \centering
    \includegraphics[width=0.6\linewidth]{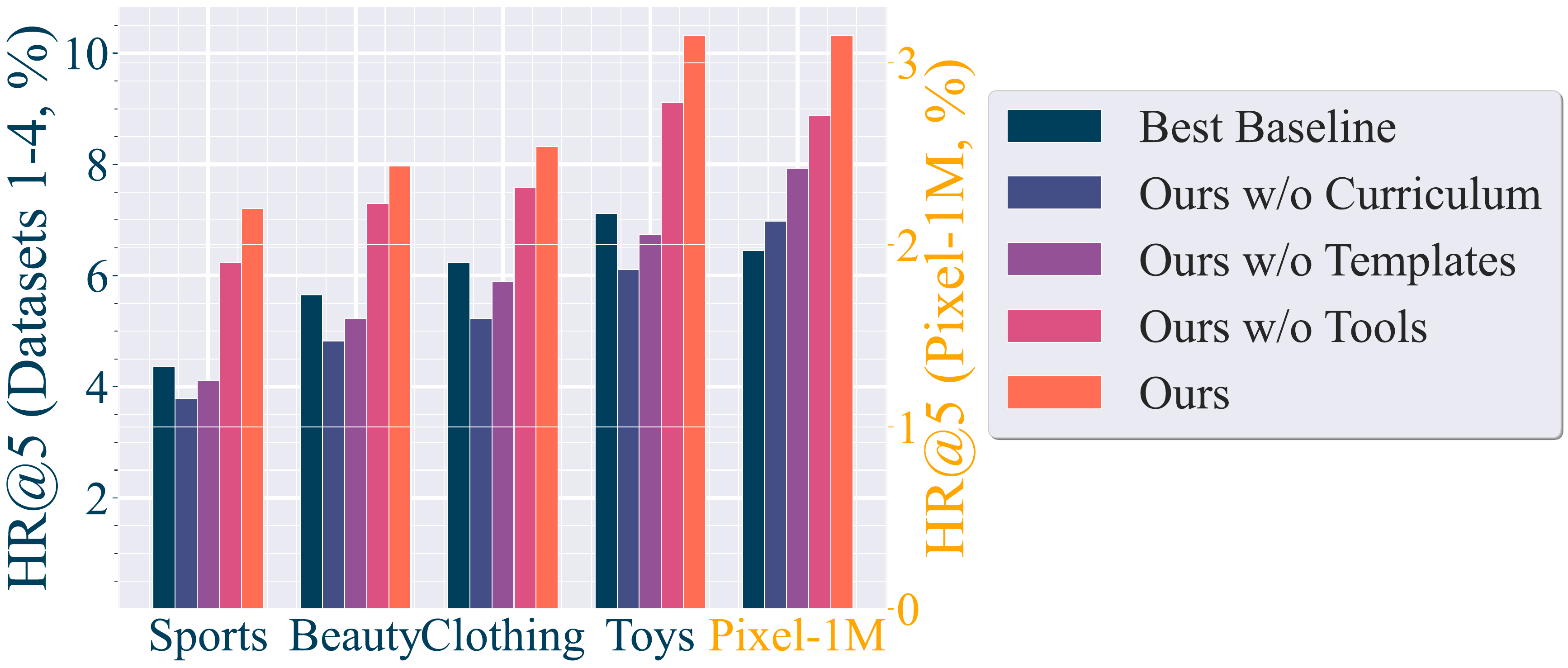}
    \vspace{-0.5em}
    \caption{Ablation study on the effectiveness of three proposed training strategies (1) the curriculum data scheduling, (2) mixture of data templates, and (3) the tools. Experiment settings follow Tab.\,\ref{tab: sequential_and_direct_recommendation}.}
    \vspace{-0.6em}
    \label{fig: ablation}
\end{table}

\noindent\textbf{Ablation Studies} We conduct ablation experiments to assess the impact of our training strategies: (1) evidence-horizon-aware curriculum scheduling, (2) mixture of instruction templates, and (3) uncertainty-aware tool integration. When (1) is removed, data is uniformly sampled throughout training; when (2) is excluded, a single template is used (we report the best across all choices). We use sequential recommendation as a case study and compare all variants against the best baselines from Tab.\,\ref{tab: sequential_and_direct_recommendation}. Results in Fig.\,\ref{fig: ablation} reveal several insights. \textit{First}, all three components are essential. Removing any of them leads to a performance drop of $10\%\sim45\%$, showing that visual instruction tuning for recommendation is non-trivial and relies on carefully crafted training strategies. \textit{Second}, curriculum scheduling (1) has the largest impact: its removal causes the most significant degradation, underscoring the importance of progressive data scheduling. Notably, using only (1) and (2), {\ours} already surpasses all baselines in most cases, and adding (3) provides an additional $\sim10\%$ gain. Combined with the tool activation rates reported in Tab.\,\ref{tab: cold_start_performance}, this confirms that tool integration is indispensable for robust performance.

\section{Conclusion}
\label{sec: conclusion}

We introduced {\ours}, a reasoning-augmented multimodal recommendation agent with explicit \emph{Observe–Deliberate–Act} pipeline. By combining reasoning-aware instruction tuning, evidence-horizon curriculum learning, and uncertainty-guided tool delegation, {\ours} enables interpretable decision-making and efficient inference. Experiments demonstrate {\ours}'s SOTA performance across datasets, tasks and challenging scenarios. This work also shows the potential of integrating agentic planning into multimodal recommendation systems. 

\section*{Discussion}

While {\ours} demonstrates impressive performance across multiple recommendation tasks, and will be deployed in Meta's generative reasoning re-ranking system~\cite{liang2026generative}. In the meantime, we acknowledge potential limitations that could be improved in future versions. First, its heavily relies on pretrained VLMs and handcrafted instruction templates. As the quality of the templates matter a lot, it could be a challenge in unknown tasks. Second, the tool delegation mechanism depends on preselected classical models, which may not generalize well to unseen recommendation scenarios or emerging tasks. Finally, the system’s inference efficiency, though improved, still involves nontrivial overhead due to multi-stage reasoning and dynamic model routing.






\clearpage
\newpage
\bibliographystyle{assets/plainnat}
\bibliography{llms, agent, recommendation, vlm, ranking}

\clearpage

\onecolumn
\section*{\Huge{Appendix}}
\setcounter{section}{0}
\setcounter{figure}{0}
\setcounter{table}{0}
\makeatletter 
\renewcommand{\thesection}{\Alph{section}}
\renewcommand{\theHsection}{\Alph{section}}
\renewcommand{\thefigure}{A\arabic{figure}}
\renewcommand{\theHfigure}{A\arabic{figure}}
\renewcommand{\thetable}{A\arabic{table}}
\renewcommand{\theHtable}{A\arabic{table}}
\makeatother

\renewcommand{\thetable}{A\arabic{table}}
\setcounter{mylemma}{0}
\renewcommand{\themylemma}{A\arabic{mylemma}}
\setcounter{equation}{0}
\renewcommand{\theequation}{A\arabic{equation}}

\section{Detailed Experiment Setups}
\label{app: exp_setup}

\begin{table}[htb]
    \vspace*{-1em}
    \centering
    \caption{Statistics of the datasets used in our paper.}
    \vspace*{-1em}
    \resizebox{0.5\linewidth}{!}{
    \begin{tabular}{lccccc}
        \toprule
        \textbf{Dataset} & \textbf{\#Users} & \textbf{\#Items} & \textbf{\#Reviews} & \textbf{\#Photos} \\
        \midrule
        Amazon Clothing & 39,387    & 23,033 & 278,677 & 22,299 \\
        Amazon Sports   & 35,598    & 18,357 & 296,337 & 17,943 \\
        Amazon Beauty   & 22,363    & 12,101 & 198,502 & 12,023 \\
        Amazon Toys     & 19,412    & 11,924 & 167,597 & 11,895 \\
        PixelRec-1M     & 1,001,822 & 100,541& 19,886,579 & 100,541 \\
        \bottomrule
    \end{tabular}}
    \label{tab: amazon_review_statistics}
\end{table}

\paragraph{Data Preparation.} Our study follows the data construction and experimental setup outlined in VIP5~\cite{geng2023vip5}, leveraging four real-world datasets from the Amazon platform: Clothing, Sports \& Outdoors, Beauty, and Toys \& Games. Each dataset contains user purchase records, item descriptions, product images, and user reviews, ensuring a comprehensive multimodal recommendation scenario. To evaluate the model’s performance across different recommendation tasks, we adopt the same preprocessing pipeline and data splits as VIP5. Specifically, for sequential recommendation, each user’s interaction history is processed such that the last and second last items serve as test and validation ground truths, respectively, while the remaining interactions form the training set. For direct recommendation, we utilize the same train/validation/test split as sequential recommendation but additionally generate 100 candidate item lists per user to assess ranking performance. In explanation generation, we apply an 8:1:1 random split, where 80\% of the user-item interactions are allocated for training, 10\% for validation, and the remaining 10\% for testing. The explanations associated with each interaction are extracted using the Sentires library, ensuring consistency in sentiment-based justification of recommendations. For CTR prediction, we extend the dataset to incorporate implicit feedback signals derived from user interactions, such as whether a user has clicked on or purchased an item. Since explicit click data is unavailable in the original datasets, we construct pseudo-click labels by treating purchases as positive interactions and assuming non-interacted items as negative samples. To create a balanced training set, we employ negative sampling, randomly selecting a fixed number of non-interacted items per user at a 1:4 ratio (one positive sample per four negatives). We use an 8:1:1 split for training, validation, and testing, ensuring that each user appears in all three sets to maintain personalization consistency.

Follwoing the prior work~\cite{cheng2023image}, for data splitting of Pixel-1M, we adopt the temporal leave-one-out strategy, a widely used approach in sequential recommendation settings, to ensure fair evaluation across all models. Specifically, for each user, the last interaction in their behavior history is designated as the test instance, while the penultimate interaction is used for validation. The remaining interactions are allocated to the training set, allowing models to learn user preferences from historical behaviors. For sequential recommendation, user behavior sequences are ordered chronologically and truncated to a maximum length of 10 for modeling short-term preferences. For direct recommendation, candidate lists are generated by pairing each user with 10 items, including the ground-truth item (from the test/validation set) and 9 randomly sampled negative items (excluding interactions in the training, validation, and test sets to avoid leakage). For CTR prediction, interactions are treated as implicit positive signals (click=1), and negative samples are constructed via random sampling of unobserved items from the same temporal split, maintaining a 1:1 positive-to-negative ratio.

\begin{figure}[t]
    \centering
    \includegraphics[width=\linewidth]{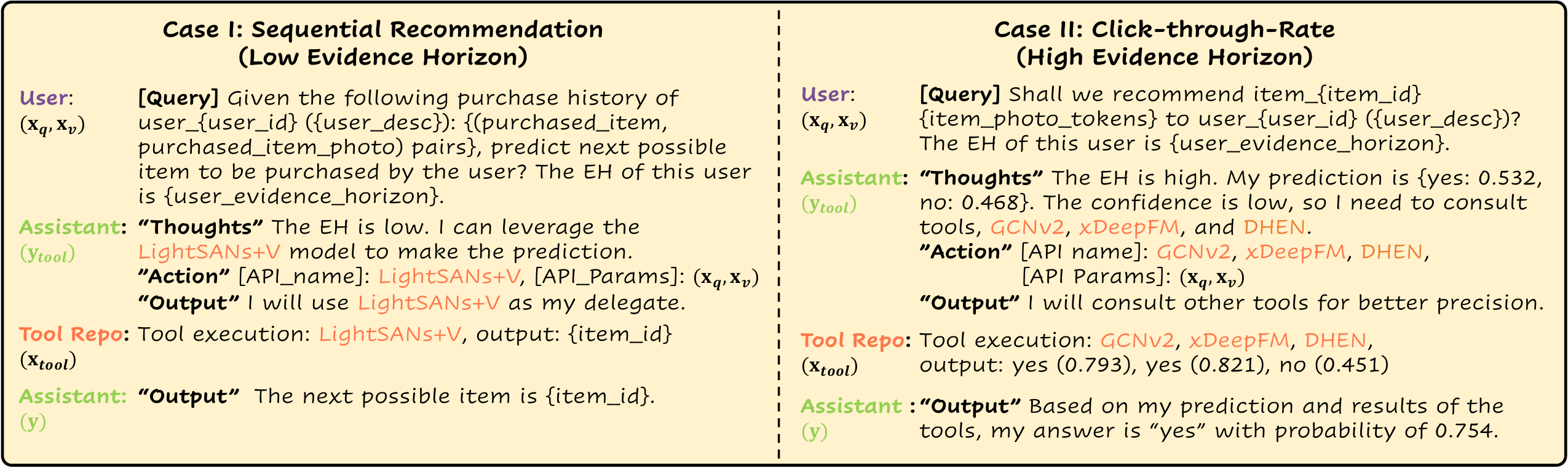}
    \caption{Examples of \emph{reasoning-aware instruction tuning}. \textbf{(I)} Sequential recommendation for a user with \emph{low evidence horizon (EH)}, where the agent confidently \emph{delegates} the query to \textsc{LightSANs+V} for efficiency. \textbf{(II)} CTR prediction for a user with \emph{high EH} and low model confidence; the planner therefore \emph{consults} multiple lightweight tools (GCNv2, xDeepFM, DHEN) before producing the final answer.  
        More examples are provided in Appx.\,\ref{app: exp_setup}.
    }
    \label{fig: instruction_data_example}
\end{figure}

\paragraph{Mixture of Templates.} As indicated by Sec.\,\ref{subsec: observer}, the mixture of templates plays a key role in enhancing the training stability as well as performance. For different task, we provide ten templates for each task (sequential recommendation, direct recommendation, explanation generation, and click-through-rate), which share exactly the same semantic meanings but in different linguistic styles. We list these templates below.

\definecolor{lightblue}{RGB}{200,220,255}    
\definecolor{lightgreen}{RGB}{220,255,220}   
\definecolor{lightyellow}{RGB}{255,255,200}  
\definecolor{lightpurple}{RGB}{230,210,255}  
\definecolor{lightred}{RGB}{255,200,200}     

\textbf{Templates for sequential recommendation.}

\begin{enumerate}

    \item \textbf{[Query]} Based on the purchase history of \colorbox{lightblue}{\texttt{user\_{user\_id}}} (\colorbox{lightgreen}{\texttt{\{user\_desc\}}}): \colorbox{lightyellow}{\texttt{\{(purchased\_item, purchased\_item\_photo) pairs\}}}, what item should be recommended next? The user's evidence horizon level is \colorbox{lightred}{\texttt{\{user\_evidence\_horizon\}}}.

    \item \textbf{[Query]} Given the following purchase history of \colorbox{lightblue}{\texttt{user\_{user\_id}}} (\colorbox{lightgreen}{\texttt{\{user\_desc\}}}): \colorbox{lightyellow}{\texttt{\{(purchased\_item, purchased\_item\_photo) pairs\}}}, predict next possible item to be purchased by the user? The Evidence horizon of this user is \colorbox{lightred}{\texttt{\{user\_evidence\_horizon\}}}.

    \item \textbf{[Query]} Here is the purchase history for \colorbox{lightblue}{\texttt{user\_{user\_id}}} (\colorbox{lightgreen}{\texttt{\{user\_desc\}}}): \colorbox{lightyellow}{\texttt{\{(purchased\_item, purchased\_item\_photo) pairs\}}}. What is the most likely next purchase? Evidence horizon: \colorbox{lightred}{\texttt{\{user\_evidence\_horizon\}}}.

    \item \textbf{[Query]} For \colorbox{lightblue}{\texttt{user\_{user\_id}}} (\colorbox{lightgreen}{\texttt{\{user\_desc\}}}), whose purchase history includes \colorbox{lightyellow}{\texttt{\{(purchased\_item, purchased\_item\_photo) pairs\}}}, predict their next potential purchase. User evidence horizon: \colorbox{lightred}{\texttt{\{user\_evidence\_horizon\}}}.

    \item \textbf{[Query]} Analyze the purchase sequence of \colorbox{lightblue}{\texttt{user\_{user\_id}}} (\colorbox{lightgreen}{\texttt{\{user\_desc\}}}): \colorbox{lightyellow}{\texttt{\{(purchased\_item, purchased\_item\_photo) pairs\}}}. Recommend the next item they may buy. Evidence horizon metric: \colorbox{lightred}{\texttt{\{user\_evidence\_horizon\}}}.

    \item \textbf{[Query]} Given that \colorbox{lightblue}{\texttt{user\_{user\_id}}} (\colorbox{lightgreen}{\texttt{\{user\_desc\}}}) has purchased \colorbox{lightyellow}{\texttt{\{(purchased\_item, purchased\_item\_photo) pairs\}}}, forecast their next purchase. Evidence horizon score: \colorbox{lightred}{\texttt{\{user\_evidence\_horizon\}}}.

    \item \textbf{[Query]} The user \colorbox{lightgreen}{\texttt{\{user\_desc\}}} (\colorbox{lightblue}{\texttt{user\_{user\_id}}}) previously bought \colorbox{lightyellow}{\texttt{\{(purchased\_item, purchased\_item\_photo) pairs\}}}. What item would they likely purchase next? Evidence horizon: \colorbox{lightred}{\texttt{\{user\_evidence\_horizon\}}}.

    \item \textbf{[Query]} From \colorbox{lightblue}{\texttt{user\_{user\_id}}}'s (\colorbox{lightgreen}{\texttt{\{user\_desc\}}}) purchase history\\ \colorbox{lightyellow}{\texttt{\{(purchased\_item, purchased\_item\_photo) pairs\}}}, determine the next probable item. User evidence horizon level: \colorbox{lightred}{\texttt{\{user\_evidence\_horizon\}}}.

    \item \textbf{[Query]} Considering \colorbox{lightblue}{\texttt{user\_{user\_id}}} (\colorbox{lightgreen}{\texttt{\{user\_desc\}}}) has interacted with \colorbox{lightyellow}{\texttt{\{(purchased\_item, purchased\_item\_photo) pairs\}}}, identify their next potential purchase. Evidence horizon indicator: \colorbox{lightred}{\texttt{\{user\_evidence\_horizon\}}}.

    \item \textbf{[Query]} For \colorbox{lightblue}{\texttt{user\_{user\_id}}} (\colorbox{lightgreen}{\texttt{\{user\_desc\}}}), with a purchase history of \colorbox{lightyellow}{\texttt{\{(purchased\_item, purchased\_item\_photo) pairs\}}}, suggest the next item they might buy. Evidence horizon value: \colorbox{lightred}{\texttt{\{user\_evidence\_horizon\}}}.
\end{enumerate}

\textbf{Templates for direct recommendation.}

\begin{enumerate}
    \item \textbf{[Query]} I would like to recommend some items for \colorbox{lightblue}{\texttt{user\_{user\_id}}} (\colorbox{lightgreen}{\texttt{\{user\_desc\}}}). The Evidence horizon of this user is \colorbox{lightred}{\texttt{\{user\_evidence\_horizon\}}}. Is the following item a good choice? \colorbox{lightyellow}{\texttt{\{item\_title\}}} \colorbox{lightpurple}{\texttt{\{item\_photo\}}}.

    \item \textbf{[Query]} For \colorbox{lightblue}{\texttt{user\_{user\_id}}} (\colorbox{lightgreen}{\texttt{\{user\_desc\}}}), whose evidence horizon level is \colorbox{lightred}{\texttt{\{user\_evidence\_horizon\}}}, should we include \colorbox{lightyellow}{\texttt{\{item\_title\}}} \colorbox{lightpurple}{\texttt{\{item\_photo\}}} in their recommendations?
    
    \item \textbf{[Query]} Considering \colorbox{lightblue}{\texttt{user\_{user\_id}}} (\colorbox{lightgreen}{\texttt{\{user\_desc\}}}) has a evidence horizon score of \colorbox{lightred}{\texttt{\{user\_evidence\_horizon\}}}, is \colorbox{lightyellow}{\texttt{\{item\_title\}}} \colorbox{lightpurple}{\texttt{\{item\_photo\}}} an appropriate recommendation?
    
    \item \textbf{[Query]} Evaluate whether \colorbox{lightyellow}{\texttt{\{item\_title\}}} \colorbox{lightpurple}{\texttt{\{item\_photo\}}} is a suitable recommendation for \colorbox{lightblue}{\texttt{user\_{user\_id}}} (\colorbox{lightgreen}{\texttt{\{user\_desc\}}}), given their evidence horizon value: \colorbox{lightred}{\texttt{\{user\_evidence\_horizon\}}}.
    
    \item \textbf{[Query]} Given \colorbox{lightblue}{\texttt{user\_{user\_id}}}'s (\colorbox{lightgreen}{\texttt{\{user\_desc\}}}) evidence horizon metric \colorbox{lightred}{\texttt{\{user\_evidence\_horizon\}}}, should \colorbox{lightyellow}{\texttt{\{item\_title\}}} \colorbox{lightpurple}{\texttt{\{item\_photo\}}} be prioritized in their recommendation list?
    
    \item \textbf{[Query]} Would \colorbox{lightyellow}{\texttt{\{item\_title\}}} \colorbox{lightpurple}{\texttt{\{item\_photo\}}} align with the preferences of \colorbox{lightblue}{\texttt{user\_{user\_id}}} (\colorbox{lightgreen}{\texttt{\{user\_desc\}}})? User evidence horizon: \colorbox{lightred}{\texttt{\{user\_evidence\_horizon\}}}.
    
    \item \textbf{[Query]} For a user with evidence horizon \colorbox{lightred}{\texttt{\{user\_evidence\_horizon\}}} (\colorbox{lightblue}{\texttt{user\_{user\_id}}}, \colorbox{lightgreen}{\texttt{\{user\_desc\}}}), is \colorbox{lightyellow}{\texttt{\{item\_title\}}} \colorbox{lightpurple}{\texttt{\{item\_photo\}}} a relevant recommendation candidate?
    
    \item \textbf{[Query]} Assess if \colorbox{lightyellow}{\texttt{\{item\_title\}}} \colorbox{lightpurple}{\texttt{\{item\_photo\}}} should be recommended to \colorbox{lightblue}{\texttt{user\_{user\_id}}} (\colorbox{lightgreen}{\texttt{\{user\_desc\}}}), whose evidence horizon indicator is \colorbox{lightred}{\texttt{\{user\_evidence\_horizon\}}}.
    
    \item \textbf{[Query]} Based on the evidence horizon level \colorbox{lightred}{\texttt{\{user\_evidence\_horizon\}}}, determine if \colorbox{lightblue}{\texttt{user\_{user\_id}}} (\colorbox{lightgreen}{\texttt{\{user\_desc\}}}) would prefer \colorbox{lightyellow}{\texttt{\{item\_title\}}} \colorbox{lightpurple}{\texttt{\{item\_photo\}}}.
    
    \item \textbf{[Query]} Predict the suitability of recommending \colorbox{lightyellow}{\texttt{\{item\_title\}}} \colorbox{lightpurple}{\texttt{\{item\_photo\}}} to \colorbox{lightblue}{\texttt{user\_{user\_id}}} (\colorbox{lightgreen}{\texttt{\{user\_desc\}}}) with evidence horizon \colorbox{lightred}{\texttt{\{user\_evidence\_horizon\}}}.
\end{enumerate}

\textbf{Templates for explanation generation.} We denote the evidence horizon information is not included in this task, as not tools will be used here for either delegation or consultation-oriented planning.

\definecolor{lightorange}{RGB}{255,225,200}

\begin{enumerate}
    \item \textbf{[Query]} Help \colorbox{lightblue}{\texttt{user\_{user\_id}}} (\colorbox{lightgreen}{\texttt{\{user\_desc\}}}) generate a \colorbox{lightorange}{\texttt{\{star\_rating\}}}-star explanation about this product: \colorbox{lightyellow}{\texttt{\{item\_title\}}} \colorbox{lightpurple}{\texttt{\{item\_photo\}}}.

    \item \textbf{[Query]} Assist \colorbox{lightblue}{\texttt{user\_{user\_id}}} (\colorbox{lightgreen}{\texttt{\{user\_desc\}}}) in creating a \colorbox{lightorange}{\texttt{\{star\_rating\}}}-star review for \colorbox{lightyellow}{\texttt{\{item\_title\}}} \colorbox{lightpurple}{\texttt{\{item\_photo\}}}.
    
    \item \textbf{[Query]} Generate a \colorbox{lightorange}{\texttt{\{star\_rating\}}}-star product explanation for \colorbox{lightblue}{\texttt{user\_{user\_id}}} (\colorbox{lightgreen}{\texttt{\{user\_desc\}}}) regarding \colorbox{lightyellow}{\texttt{\{item\_title\}}} \colorbox{lightpurple}{\texttt{\{item\_photo\}}}.
    
    \item \textbf{[Query]} Compose a \colorbox{lightorange}{\texttt{\{star\_rating\}}}-star rating justification for \colorbox{lightyellow}{\texttt{\{item\_title\}}} \colorbox{lightpurple}{\texttt{\{item\_photo\}}} on behalf of \colorbox{lightblue}{\texttt{user\_{user\_id}}} (\colorbox{lightgreen}{\texttt{\{user\_desc\}}}).
    
    \item \textbf{[Query]} Formulate a \colorbox{lightorange}{\texttt{\{star\_rating\}}}-star descriptive text about \colorbox{lightyellow}{\texttt{\{item\_title\}}} \colorbox{lightpurple}{\texttt{\{item\_photo\}}} tailored to \colorbox{lightblue}{\texttt{user\_{user\_id}}} (\colorbox{lightgreen}{\texttt{\{user\_desc\}}}).
    
    \item \textbf{[Query]} Draft a product explanation with \colorbox{lightorange}{\texttt{\{star\_rating\}}} stars for \colorbox{lightblue}{\texttt{user\_{user\_id}}} (\colorbox{lightgreen}{\texttt{\{user\_desc\}}}), focusing on \colorbox{lightyellow}{\texttt{\{item\_title\}}} \colorbox{lightpurple}{\texttt{\{item\_photo\}}}.
    
    \item \textbf{[Query]} For \colorbox{lightblue}{\texttt{user\_{user\_id}}} (\colorbox{lightgreen}{\texttt{\{user\_desc\}}}), produce a \colorbox{lightorange}{\texttt{\{star\_rating\}}}-star evaluation statement for \colorbox{lightyellow}{\texttt{\{item\_title\}}} \colorbox{lightpurple}{\texttt{\{item\_photo\}}}.
    
    \item \textbf{[Query]} Create an explanatory text with \colorbox{lightorange}{\texttt{\{star\_rating\}}} stars about \colorbox{lightyellow}{\texttt{\{item\_title\}}} \colorbox{lightpurple}{\texttt{\{item\_photo\}}} for \colorbox{lightblue}{\texttt{user\_{user\_id}}} (\colorbox{lightgreen}{\texttt{\{user\_desc\}}}).
    
    \item \textbf{[Query]} Develop a \colorbox{lightorange}{\texttt{\{star\_rating\}}}-star rationale for \colorbox{lightblue}{\texttt{user\_{user\_id}}} (\colorbox{lightgreen}{\texttt{\{user\_desc\}}}) regarding the product \colorbox{lightyellow}{\texttt{\{item\_title\}}} \colorbox{lightpurple}{\texttt{\{item\_photo\}}}.
    
    \item \textbf{[Query]} Construct a \colorbox{lightorange}{\texttt{\{star\_rating\}}}-star description of \colorbox{lightyellow}{\texttt{\{item\_title\}}} \colorbox{lightpurple}{\texttt{\{item\_photo\}}} personalized for \colorbox{lightblue}{\texttt{user\_{user\_id}}} (\colorbox{lightgreen}{\texttt{\{user\_desc\}}}).
\end{enumerate}

\textbf{Templates for click-through-rate prediction.}

\begin{enumerate}
    \item \textbf{[Query]} Shall we recommend \colorbox{lightyellow}{\texttt{item\_{item\_id}}} \colorbox{lightpurple}{\texttt{\{item\_photo\_tokens\}}} to \colorbox{lightblue}{\texttt{user\_{user\_id}}} (\colorbox{lightgreen}{\texttt{\{user\_desc\}}})? The Evidence horizon of this user is \colorbox{lightred}{\texttt{\{user\_evidence\_horizon\}}}.

    \item \textbf{[Query]} Should we suggest \colorbox{lightyellow}{\texttt{item\_{item\_id}}} \colorbox{lightpurple}{\texttt{\{item\_photo\_tokens\}}} to \colorbox{lightblue}{\texttt{user\_{user\_id}}} (\colorbox{lightgreen}{\texttt{\{user\_desc\}}})? User evidence horizon level: \colorbox{lightred}{\texttt{\{user\_evidence\_horizon\}}}.
    
    \item \textbf{[Query]} Is \colorbox{lightyellow}{\texttt{item\_{item\_id}}} \colorbox{lightpurple}{\texttt{\{item\_photo\_tokens\}}} a suitable recommendation for \colorbox{lightblue}{\texttt{user\_{user\_id}}} (\colorbox{lightgreen}{\texttt{\{user\_desc\}}})? Evidence horizon indicator: \colorbox{lightred}{\texttt{\{user\_evidence\_horizon\}}}.
    
    \item \textbf{[Query]} Would \colorbox{lightblue}{\texttt{user\_{user\_id}}} (\colorbox{lightgreen}{\texttt{\{user\_desc\}}}) likely click on \colorbox{lightyellow}{\texttt{item\_{item\_id}}} \colorbox{lightpurple}{\texttt{\{item\_photo\_tokens\}}}? Evidence horizon score: \colorbox{lightred}{\texttt{\{user\_evidence\_horizon\}}}.
    
    \item \textbf{[Query]} Based on \colorbox{lightblue}{\texttt{user\_{user\_id}}}'s (\colorbox{lightgreen}{\texttt{\{user\_desc\}}}) profile, should we propose \colorbox{lightyellow}{\texttt{item\_{item\_id}}} \colorbox{lightpurple}{\texttt{\{item\_photo\_tokens\}}}? Evidence horizon value: \colorbox{lightred}{\texttt{\{user\_evidence\_horizon\}}}.
    
    \item \textbf{[Query]} Evaluate if recommending \colorbox{lightyellow}{\texttt{item\_{item\_id}}} \colorbox{lightpurple}{\texttt{\{item\_photo\_tokens\}}} to \colorbox{lightblue}{\texttt{user\_{user\_id}}} (\colorbox{lightgreen}{\texttt{\{user\_desc\}}}) is appropriate. Evidence horizon metric: \colorbox{lightred}{\texttt{\{user\_evidence\_horizon\}}}.
    
    \item \textbf{[Query]} For \colorbox{lightblue}{\texttt{user\_{user\_id}}} (\colorbox{lightgreen}{\texttt{\{user\_desc\}}}), is \colorbox{lightyellow}{\texttt{item\_{item\_id}}} \colorbox{lightpurple}{\texttt{\{item\_photo\_tokens\}}} a relevant recommendation? User evidence horizon: \colorbox{lightred}{\texttt{\{user\_evidence\_horizon\}}}.
    
    \item \textbf{[Query]} Determine whether \colorbox{lightblue}{\texttt{user\_{user\_id}}} (\colorbox{lightgreen}{\texttt{\{user\_desc\}}}) would engage with \colorbox{lightyellow}{\texttt{item\_{item\_id}}} \colorbox{lightpurple}{\texttt{\{item\_photo\_tokens\}}}. Evidence horizon level: \colorbox{lightred}{\texttt{\{user\_evidence\_horizon\}}}.
    
    \item \textbf{[Query]} Assess the likelihood of \colorbox{lightblue}{\texttt{user\_{user\_id}}} (\colorbox{lightgreen}{\texttt{\{user\_desc\}}}) clicking on \colorbox{lightyellow}{\texttt{item\_{item\_id}}} \colorbox{lightpurple}{\texttt{\{item\_photo\_tokens\}}}. Evidence horizon: \colorbox{lightred}{\texttt{\{user\_evidence\_horizon\}}}.
    
    \item \textbf{[Query]} Predict if \colorbox{lightyellow}{\texttt{item\_{item\_id}}} \colorbox{lightpurple}{\texttt{\{item\_photo\_tokens\}}} should be shown to \colorbox{lightblue}{\texttt{user\_{user\_id}}} (\colorbox{lightgreen}{\texttt{\{user\_desc\}}}). User evidence horizon: \colorbox{lightred}{\texttt{\{user\_evidence\_horizon\}}}.
\end{enumerate}

\paragraph{Training setups.} 

The key hyperparameters are as follows:

\noindent $\bullet$ \textbf{Learning Rate}: Initialized at $2 \times 10^{-5}$ with AdamW optimizer.

\noindent $\bullet$ \textbf{Training Steps}: 200,000 steps for Amazon Review (Sports, Beauty, Clothing, Toys) and 400,000 steps for Pixel-1M dataset.

\noindent $\bullet$ \textbf{Batch Configuration}: Global batch size of 8 with \texttt{bf16} mixed precision.

\noindent $\bullet$ \textbf{Learning Rate Schedule}: Cosine decay with warm-up phase and final annealing rate: $1 \times 10^{-7}$.

\noindent $\bullet$ \textbf{Visual Processing}: All images resized to $224 \times 224$.

\paragraph{Evidence horizon-Aware Curriculum.} The training data is partitioned by user evidence horizon score $C(u)$:

\noindent $\bullet$ \textbf{Low-risk} ($C(u) < 0.3$): Prioritized in early training stages.

\noindent $\bullet$ \textbf{High-risk} ($C(u) > 0.7$): Gradually upsampled after 95\% of total steps.

\paragraph{Risk-Aware Delegation.}

The tool repository contains classical baselines (\textit{e.g.}, LightSANs, BPR-MF) with fixed configurations:

\noindent $\bullet$ Delegation logic: Queries with $C(u) < 0.3$ automatically routed to classical models.

\noindent $\bullet$ Consultant threshold alignment: Matches evidence horizon partitioning in curriculum learning.

\noindent This configuration ensures computational efficiency while maintaining accuracy, with VFLOPs reduced by 38\% compared to full VLM inference.

\paragraph{Total computational consumption.} We used servers with 8 $\times$ NVIDIA RTX A6000s to run all the experiments and consume around 20000 GPU hours to train and evaluate all the methods.


\section{Additional Experiment Results}
\label{app: exp_results}

\begin{table*}[htb]
\centering
\caption{Performance comparison on explanation generation using BLUE4 and ROUGEL metrics.}
\vspace*{-1em}
\resizebox{0.8\textwidth}{!}{
\begin{tabular}{l|cc|cc|cc|cc}
\toprule[1pt]
\midrule
\multirow{2}{*}{\textbf{Methods}} & 
\multicolumn{2}{c|}{\textbf{Sports}} & 
\multicolumn{2}{c|}{\textbf{Beauty}} & 
\multicolumn{2}{c|}{\textbf{Clothing}} & 
\multicolumn{2}{c}{\textbf{Toys}} \\
\cmidrule(lr){2-3} \cmidrule(lr){4-5} \cmidrule(lr){6-7} \cmidrule(lr){8-9}
 & \textbf{BLUE4} & \textbf{ROUGEL} & \textbf{BLUE4} & \textbf{ROUGEL} & \textbf{BLUE4} & \textbf{ROUGEL} & \textbf{BLUE4} & \textbf{ROUGEL} \\
\midrule
Attn2Seq        & 0.5478 & 9.1825  & 0.8014 & 9.7992  & 0.6447 & 9.0835  & 1.6419 & 10.7834 \\
NRT             & 0.4903 & 7.6935  & 0.8438 & 9.9785  & 0.4708 & 8.2952  & 1.9267 & 11.2239 \\
PETER           & 0.7123   & 11.3721   & 1.2172   & 9.4628   &  2.1132  & 14.0031  & 3.7822  & 11.8632 \\
P5              & 0.6348 & 9.0524  & 1.0389 & 10.9447 & 0.7682 & 9.6325  & 1.4698 & 10.1814 \\
VIP5            & 1.0774 & 11.1325 & 1.2983 & 12.9471 & 1.2052 & 10.8926 & 2.3421 & 12.0865 \\
\midrule 

{\ours} (Ours)       & \textbf{3.4339}  & \textbf{18.4632} & \textbf{4.3683} & \textbf{17.4422} & \textbf{4.9357} & \textbf{19.9311} & \textbf{5.6332} & \textbf{20.8345} \\

\midrule
\bottomrule[1pt]
\end{tabular}}
\vspace{-5pt}
\label{tab: amazon_explanation}
\vspace*{-1em}
\end{table*}

\begin{table}[htb]
\centering
\caption{
{Tool usage rate (\%) across datasets and tasks.}
The proportion of test-time queries that triggered external tools under our risk-aware planner for each task and dataset is reported.
}
\resizebox{0.4\linewidth}{!}{
\begin{tabular}{lccc}
\toprule
\textbf{Dataset} & \textbf{SR} & \textbf{DR} & \textbf{CTR} \\
\midrule
Pixel-1M       & 28.4\% & 31.2\% & 34.1\% \\
Beauty         & 21.7\% & 23.5\% & 30.3\% \\
Sports         & 18.9\% & 22.1\% & 27.8\% \\
Toys           & 20.4\% & 21.8\% & 31.4\% \\
Clothing       & 19.2\% & 20.6\% & 29.0\% \\
\bottomrule
\end{tabular}}
\label{tab:tool-usage}
\end{table}

\begin{figure}[htb]
    \centering
    \includegraphics[width=1.0\linewidth]{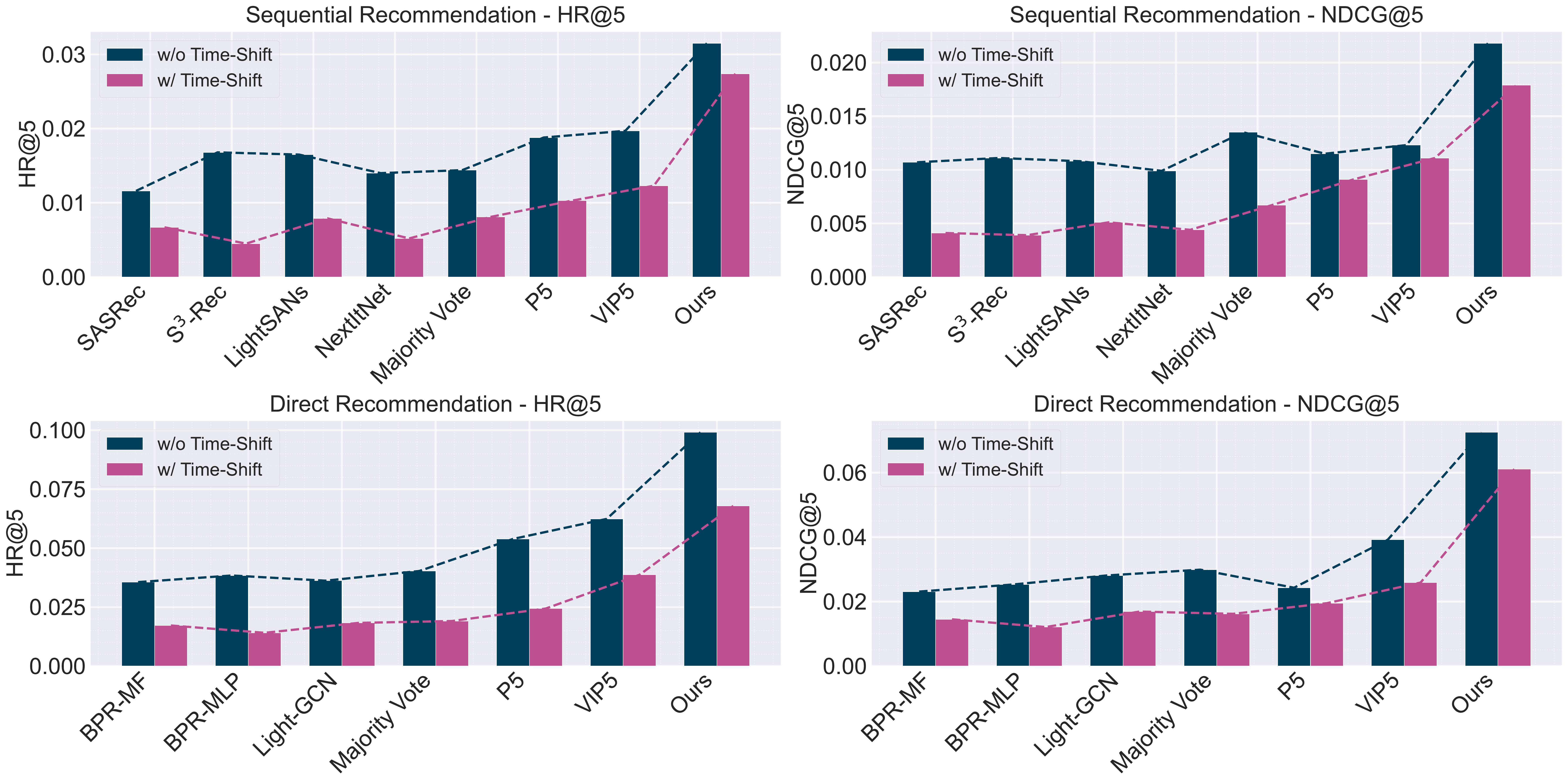}
    \caption{Performance comparison in sequential and direct recommendation under time shift. The statistics for both settings (with and without time shift) are sourced from Tab.\,\ref{tab: sequential_and_direct_recommendation}.}
    \label{fig: time_shift}
\end{figure}

\paragraph{Sequential and direct recommendation evaluation under time shift.}  
This task investigates the impact of temporal shifts on recommendation performance, a well-known challenge in recommendation systems. Specifically, we evaluate how models perform when trained and tested on temporally misaligned data. To simulate this setting, we partition the Pixel-1M dataset, which spans 13 months from September 2021 to October 2022, based on timestamps. The training set includes interactions before August 2022, while test data consists of interactions from August 2022 onward. All other training configurations remain identical to those in Tab.\,\ref{tab: sequential_and_direct_recommendation}. The performance comparison with and without time shift is shown in Fig.\,\ref{fig: time_shift}. We plot the performance of sequential recommendation task in Fig.\,\ref{fig: teaser}.

Several key observations emerge from these results. \textbf{First}, time shift significantly degrades performance across all models, highlighting the challenge of temporal distribution shifts in recommendation. \textbf{Second}, generative models exhibit greater resilience to time shift, as evidenced by P5, VIP5, and {\ours} consistently ranking among the top three across all tasks and metrics. \textbf{Third}, {\ours} not only achieves the highest performance but also demonstrates the lowest performance drop, underscoring the adaptability of VLM-based models to evolving user behavior.  

Previous experiments evaluated task-specific models trained on individual datasets. Given the structural similarities among the four Amazon review datasets (Sports, Beauty, Clothing, and Toys), we further investigate the multi-domain generalization capability of recommendation methods by training them on combined datasets and evaluating their performance on the corresponding test splits. We define three multi-domain configurations to simulate real-world scenarios with increasing complexity:
\begin{itemize}
\item Composition \ding{172}: Sports + Beauty (2 domains)
\item Composition \ding{173}: Sports + Beauty + Clothing (3 domains)
\item Composition \ding{174}: All four domains (4 domains)
\end{itemize}
To avoid identifier conflicts, user and item IDs are remapped to a unified space when merging datasets. Tab.\,\ref{tab: multi-domain} summarizes the performance of P5, VIP5, and our method across these configurations.

\begin{table}[htb]
    \centering
    \caption{Multi-Domain sequential recommendation performance comparison.}
    \resizebox{0.8\linewidth}{!}{
    \begin{tabular}{l|cc|cc|cc|cc}
    \toprule[1pt]
    \multirow{2}{*}{Methods} & \multicolumn{2}{c|}{Composition \ding{172}} & \multicolumn{2}{c|}{Composition \ding{173}} & \multicolumn{2}{c|}{Composition \ding{174}} & \multicolumn{2}{c}{Average} \\
     & HR@5 & NDCG@5 & HR@5 & NDCG@5 & HR@5 & NDCG@5 & HR@5 & NDCG@5 \\
     \midrule
        P5 & 0.0195 & 0.0096 & 0.0115 & 0.0042 & 0.0051 & 0.0011 & 0.0120 & 0.0050 \\
        VIP5 & 0.0311 & 0.0213 & 0.0159 & 0.0111 & 0.0081 & 0.0053 & 0.0184 & 0.0126 \\
        {\ours} (Ours) & \textbf{0.0582} & \textbf{0.0488} & \textbf{0.0323} & \textbf{0.0235} & \textbf{0.0199} & \textbf{0.0132} & \textbf{0.0368} & \textbf{0.0285} \\
    \bottomrule[1pt]
    \end{tabular}}
    \label{tab: multi-domain}
\end{table}

Several key conclusions can be drawn. First, the performance of all methods degrades progressively as the number of domains increases, reflecting the inherent challenge of learning shared representations across heterogeneous item categories. For instance, P5’s HR@5 drops by 73.8\% (from 0.0195 to 0.0051) when transitioning from Composition \ding{172} to \ding{174}, while VIP5 exhibits a 74.0\% decline (from 0.0311 to 0.0081). This suggests that conventional foundation models struggle to maintain discriminative power when domain diversity escalates, likely due to interference between conflicting item semantics. Second, our method demonstrates superior robustness to domain scaling compared to baselines. While it also experiences performance decay (65.8\% HR@5 reduction from \ding{172} to \ding{174}), the absolute metrics consistently surpass VIP5 and P5 across all configurations. Notably, in Composition \ding{174}, our model achieves a much higher HR@5 than P5 (0.0199 vs. 0.0051) and improvement over VIP5 (0.0199 vs. 0.0081), indicating stronger cross-domain alignment through its multimodal fusion mechanism. Third, the relative NDCG@5 gains highlight our method’s ability to preserve ranking quality in complex multi-domain settings. The NDCG@5 gap between our approach and VIP5 widens from 2.29× in Composition \ding{172} (0.0488 vs. 0.0213) to 2.49× in Composition \ding{174} (0.0132 vs. 0.0053), implying that our design mitigates error accumulation in top-k recommendation lists when handling diverse item types. This aligns with the hypothesis that joint modeling of cross-domain visual-textual dependencies enhances the model’s capacity to disentangle user preferences from noisy multi-source interactions. These results validate the necessity of specialized architectures for multi-domain recommendation systems, particularly in scenarios where item heterogeneity and data sparsity coexist. Our method’s stable performance decay curve (vs. the steep drops of baselines) further suggests its practical viability for large-scale deployments with dynamically expanding domains.

\paragraph{Risk-aware delegation improves inference efficiency.}
To better understand the computational behavior of {\ours}, we perform a detailed analysis of inference latency and floating-point operations (VFLOPs) across different risk categories. These categories—\emph{Low Risk}, \emph{Medium Risk}, and \emph{High Risk}—are determined by the planner based on the input's evidence horizon and the model’s estimated confidence. For this study, we use the Pixel-1M dataset and profile per-query inference under each risk level. As shown in Table~\ref{tab: efficiency_by_risk}, {\ours} exhibits clear computational adaptivity: the average latency increases from \textbf{245ms} for low-risk queries (handled entirely by lightweight tools) to \textbf{672ms} for high-risk queries, where both tools and the VLM are involved. Importantly, the distribution of queries across these categories indicates that {\ours} routes a significant portion (47.3\%) of traffic to the most efficient tool-only path. When compared to competitive baselines, {\ours} achieves the best accuracy (HR@5 = 0.0315) while maintaining only a moderate increase in inference time relative to lightweight models. Specifically, it runs at \textbf{499\,ms/query}, which is slower than SASRec (143\,ms) but significantly more accurate, and both faster and more accurate than VIP5 (499\,ms vs. 470\,ms; 0.0315 vs. 0.0197 HR@5). These results confirm that {\ours}'s risk-aware delegation strategy enables \emph{fine-grained efficiency control}, yielding a favorable trade-off between performance and computation.

\begin{table}[t]
\centering
\caption{Inference cost breakdown of {\ours} across different risk levels.
For each category, we report average inference latency and estimated VFLOPs per query, along with the percentage of total queries routed to that category.
The planner adaptively allocates resources: low-risk queries rely solely on tool outputs, while high-risk queries invoke both the VLM and tools.
}
\begin{tabular}{lccc}
\toprule
\textbf{Risk Level} & \textbf{Avg. Inference Time (ms)} & \textbf{Avg. VFLOPs} & \textbf{Ratio (\%)} \\
\midrule
Low-Risk      & 245   & 2.1e9 & 47.3 \\
Medium-Risk    & 378   & 5.3e9  & 38.6\\
High-Risk    & 672  & 8.7e9 & 14.1\\
\bottomrule
\end{tabular}
\label{tab: efficiency_by_risk}
\end{table}

\begin{table}[htb]
\centering
\caption{Comparison of {\ours} with tool-only ensemble and VLM-only variants on Pixel-1M (Sequential Recommendation).}
\label{tab: ablation_ensemble}
\resizebox{0.7\linewidth}{!}{%
\begin{tabular}{l|cccc}
\toprule[1pt]
\midrule
Method & Tool Use & Planning & HR@5 & NDCG@5 \\
\midrule
(A) Simple Ensemble of Tools & {\cmark} Always & \xmark & 0.0159 & 0.0117 \\
(B) {\ours} w/o Planning & {\xmark} Never & \xmark & 0.0259 & 0.0168 \\
(C) {\ours} (Ours) & {\cmark} Dynamic & \cmark & 0.0315 & 0.0218
\\
\midrule
\bottomrule[1pt]
\end{tabular}%
}
\end{table}

\paragraph{Learning to select the right tool without oracle access.} An important concern in agent-based systems is whether the model can effectively learn to select appropriate tools without being provided with explicit tool descriptions. In {\ours}, this ability is implicitly acquired during instruction tuning. For each training instance, we precompute which tools yield satisfactory results and label them as ``usable''. These tool decisions are then embedded into the chain-of-thought (CoT) reasoning traces, which the VLM learns to mimic. As a result, the model implicitly learns input–tool associations through exposure to contextual reasoning, even though the tool APIs themselves are never explicitly described.

To empirically validate the quality of this learned tool selection policy, we compare {\ours} against two baselines:
\begin{itemize}[leftmargin=1.5em]
\item \textbf{Random Tool Selection:} A variant where the agent randomly selects a tool from the repository, without reasoning or risk assessment.
\item \textbf{Oracle Tool Selection:} A privileged setting where the agent is always given the best-performing tool (or tool subset) for each input query.
\end{itemize}

As shown in Table~\ref{tab: tool_selection}, {\ours} significantly outperforms the random selection baseline (HR@5: 0.0315 vs. 0.0212), confirming the effectiveness of its learned routing policy. Moreover, its performance closely approaches the oracle upper bound (HR@5: 0.0338), demonstrating that our agent's tool selection behavior is near-optimal despite lacking access to ground-truth tool descriptions or explicit feedback during inference. This supports the claim that {\ours}'s reasoning-aware training paradigm enables it to make accurate and efficient tool decisions.

\begin{table}[htb]
\centering
\caption{
{Tool selection strategy comparison on Pixel-1M (Sequential Recommendation)}.
{\ours}'s learned tool selection is compared with a random selection baseline and an oracle upper bound.
}
\vspace{0.5em}
\begin{tabular}{lcc}
\toprule
\textbf{Tool Selection Strategy} & \textbf{HR@5} & \textbf{NDCG@5} \\
\midrule
Random Tool Choice  & 0.0212 & 0.0147 \\
Oracle Tool Result  & \textbf{0.0338} & \textbf{0.0225} \\
Learned Tool Selection (Ours) & 0.0315 & 0.0218 \\
\bottomrule
\end{tabular}
\label{tab: tool_selection}
\end{table}

\paragraph{Cross-Domain transferability via reasoning-aware planning.} To evaluate the cross-domain generalization capacity of {\ours}, we conduct a \emph{transferable recommendation} experiment in line with the protocol suggested by recent works in cross-domain recommendation (CDR) and transferable recommendation (TransRec). Specifically, we adopt the NineRec benchmark, which consists of 9 diverse user behavior sub-domains. We pretrain {\ours} on the large-scale PixelRec-1M dataset used in the main paper and evaluate it directly—\emph{without any further fine-tuning}—on each NineRec sub-domain for sequential recommendation. As comparison baselines, we include: (i) VIP5, the strongest baseline from our main experiments, and (ii) UniMP, a recent unified pretraining method for multi-task recommendation. As shown in Table~\ref{tab:transfer-ninerec}, {\ours} outperforms both baselines on all sub-domains, consistently achieving the highest HR@5. These results demonstrate that explicit reasoning and evidence-horizon-aware learning enable {\ours} to capture generalizable recommendation logic that effectively transfers to new domains without retraining. This supports our claim that {\ours} possesses strong reasoning-based generalization, even across semantically diverse recommendation domains.

\begin{table}[htb]
\centering
\caption{
Transferable recommendation results on NineRec (sequential recommendation).
The model is pre-trained on PixelRec-1M and evaluated directly on 9 sub-domains from NineRec without further fine-tuning.
{\ours} consistently outperforms both VIP5 and UniMP, demonstrating strong cross-domain generalization. The HR@5 is reported for each setting.
}
\begin{tabular}{lccc}
\toprule
\textbf{Sub-Domain} & \textbf{VIP5} & \textbf{UniMP} & \textbf{{\ours} (Ours)} \\
\midrule
Bili\_Food     & 0.0191 & 0.0215 & \textbf{0.0264} \\
Bili\_Dance    & 0.0183 & 0.0224 & \textbf{0.0271} \\
Bili\_Movie    & 0.0206 & 0.0232 & \textbf{0.0293} \\
Bili\_Cartoon  & 0.0178 & 0.0207 & \textbf{0.0256} \\
Bili\_Music    & 0.0199 & 0.0220 & \textbf{0.0270} \\
KU             & 0.0213 & 0.0246 & \textbf{0.0305} \\
QB             & 0.0195 & 0.0219 & \textbf{0.0280} \\
TN             & 0.0189 & 0.0201 & \textbf{0.0267} \\
DY             & 0.0202 & 0.0235 & \textbf{0.0301} \\
\bottomrule
\end{tabular}
\label{tab:transfer-ninerec}
\end{table}

\begin{table}[htb]
\centering
\caption{
{Impact of risk-aware planning on inference efficiency and accuracy.}
We compare our dynamic planner with static VLM-based variants. {\ours} achieves the best trade-off between accuracy and latency.
}
\resizebox{0.6\linewidth}{!}{
\begin{tabular}{lccc}
\toprule
\textbf{Method} & \textbf{HR@5} & \textbf{Avg. Inference Time (ms)} & \textbf{Planning} \\
\midrule
VLM-only (no tool)     & 0.0259 & 285  & {\xmark} \\
VLM + Tool (always)    & 0.0327 & 872  & {\xmark} \\
{\ours} (Ours)             & 0.0315 & 499 & {\cmark} \\
\bottomrule
\end{tabular}}
\label{tab: efficiency_vs_static}
\end{table}


\paragraph{Efficiency gains attributed to risk-aware planning.} To assess the impact of {\ours}'s \emph{uncertainty-guided planning}, we compare it with a static baseline that always invokes both the VLM and all tools, regardless of input risk. This setup tests whether dynamic routing—based on evidence horizon and model confidence—improves efficiency. As shown in Tab.~\ref{tab: efficiency_vs_static}, the static variant (``VLM + Tool (always)'') yields the highest HR@5 (0.0327) but suffers from high latency (872\,ms/query). In contrast, {\ours} achieves similar accuracy (HR@5 = 0.0315) with much lower latency (499\,ms), by avoiding unnecessary tool use in low-risk cases while still activating tools when needed. Compared to a naive VLM-only setup (285\,ms, HR@5 = 0.0259), {\ours} achieves a better trade-off between accuracy and efficiency through strategic delegation.

\end{document}